\documentclass[12pt,preprint]{emulateapj}

\begin{document}

\title{Applications of Fast Convolution on the Sphere}
\author{K.~M.~Huffenberger\altaffilmark{1,2}, 
I.~J.~O'Dwyer\altaffilmark{2},  
K.~M.~G\'orski\altaffilmark{2,3}, 
B.~D.~Wandelt\altaffilmark{4}
}

\altaffiltext{1}{email: huffenbe@jpl.nasa.gov}
\altaffiltext{2}{Jet Propulsion Laboratory, California Institute of Technology, Pasadena, CA  91109} 
\altaffiltext{3}{Warsaw University Observatory, Aleje Ujazdowskie 4, 00-478 Warszawa, Poland}
\altaffiltext{4}{University of Illinois at Urbana-Champaign, Urbana IL 61801} 

\begin{abstract}
We present several diverse applications of the spherical fast convolution method suggested by \citet{WG2001}, 
which is useful for studies of telescope optical properties and for construction of shaped filters for analysis of all-sky data.
We study sidelobe pickup in three design concepts for the \textit{CALISTO} infrared telescope.  The beam convolution allows for direct comparison of the performance of each telescope design.
At 100 microns, the best of these designs limits sidelobe contamination from Galactic dust emission to $< 0.1$ MJy/sr for most of the sky with $|b| > 25^\circ$. With the fast convolution method, we illustrate the impact of asymmetric primary beams on the recovery of the temperature power spectrum for the Planck microwave background mission.  Finally, we use the fast convolution method to specify a class of orientable filters on the sphere, working through a pedagogical example.   
\end{abstract}

\keywords{telescopes, methods: data analysis, infrared: general, cosmic microwave background}

\section{Introduction}
Observational astronomy and cosmology experiments require realistic simulation and analysis pipelines for detailed study of systematic effects.
In particular, the simulation of realistic telescope optical response is a computationally expensive yet important task.  Distortions of the beam (or point-spread-function) caused by imperfections in the mirrors and other optical components of the telescope lead to undesirable systematic effects, including extraneous pickup by the beam far-sidelobes.  
These effects are important and need to be modeled.  An experiment may observe a region far from the Galactic plane in order to avoid foreground contamination.  However, scattered or diffracted Galactic emission may still spoil the measurement.
At the same time, large numbers of past and planned large sky area surveys (2MASS, NVSS, SDSS, WMAP, Planck, LSST, etc.) emphasize the importance of analysis on the sphere.

In order to fully simulate and analyze these effects, it is necessary to compute the detector response in every direction: convolving a $4\pi$-steradian beam model with the sky for every possible  pointing and orientation of the telescope.  The computational cost of a brute force analysis in real space prohibits it for all but the poorest angular resolution.  Alternatively, simplifying assumptions---like beam axisymmetry---cut the cost of harmonic space calculation, but lose the orientation information, a potentially significant failing.

\citet{WG2001} devised a fast, efficient, and computationally feasible method for performing this type of all-sky convolution.  This paper highlights this algorithm with a series of applications which are difficult or impossible to address by other means.   In section \ref{sec:method}, we describe the convolution method, and the preparation of the spherical harmonic representation of the beam and sky map.  Although the applications in this work deal with total intensity only, this method has been extended to polarization by \citet{Challinor2000}.

This convolution method is used widely in two ways.  First, as described above, it can simulate telescope optical properties.  It is particularly well suited to the study of large-scale far-sidelobe pickup.  If the beam is not too small and can be described by a limited number of azimuthal harmonics, this method is also suitable for simulating the response of an asymmetric main beam. The  simulation pipeline for the upcoming \textit{Planck} Cosmic Microwave Background satellite routinely uses it for this purpose \citep{LevelS}.  In section~\ref{sec:applications}, we offer two demonstrations of fast convolution probing optical properties, describing the far-sidelobe pickup for the proposed infrared satellite mission \textit{CALISTO} and the impact of main beam asymmetries on the temperature power spectrum evaluation of Planck.

Beyond telescope optics, the second main use for this convolution method is in the analysis of all-sky astronomical signals.  For example, it has been used to look for Bianchi VIIh cosmological model contributions \citep{2005ApJ...629L...1J,2006A&A...460..393J,2006ApJ...643..616J,2006MNRAS.369..598C} to the cosmic microwave background temperature map measured by the \textit{Wilkinson Microwave Anisotropy Probe} \citep[\textit{WMAP}, e.g.][]{Hinshaw2007}.
  Also, in a number of works, the method has been used to process all-sky signals with orientable filters, in preparation for  examination of the CMB for non-Gaussian signatures, cross-correlation, or other studies
\citep{
2006MNRAS.368..226B,
2006MNRAS.369...57C,
2006ApJ...636L...1L,
2006MNRAS.369.1858M,
2007arXiv0704.3158M,
2007arXiv0704.0626M,
2006MNRAS.365..891V,
2006NewAR..50..880V,
2007arXiv0704.3736V,
2006PhRvL..96o1303W,
2007arXiv0704.3144W,
2007arXiv0706.2346W}.
Often in these works, the convolution method is used to transport around the sphere a class of filters derived from projected wavelets.  We discuss orientable filters on the sphere in section \ref{sec:applications}, 
where we emphasize that the convolution method is independent of the choice of (band-limited) filter; it does not depend on these wavelet-derived, or any other, particular filters.  To make this point, we hide a non-trivial signal under a large amount of noise, and find it with an appropriate orientable, matched filter.

Finally, discussion and conclusions follow in section \ref{sec:discussion}. 

\section{Method} \label{sec:method}

\subsection{``Total convolution''} 

The convolution problem evaluates
\begin{equation}
T(\hat \mathbf{n},\omega) = \int \,d\hat \mathbf{n}' \left[ D(\hat \mathbf{n},\omega) b \right]^{*}\!(\hat \mathbf{n}')\ s(\hat \mathbf{n}').    
\label{eqn:convolution}
\end{equation}
In this notation the vector $\hat \mathbf{n}$ defines a position on the unit sphere defined by longitude $\phi$ and colatitude $\theta$, and $\omega$ is the orientation at that position.  The rotation operator $D$ rotates the function $b(\hat \mathbf{n})$ (the ``beam'') into position from a fiducial orientation at the North pole via Euler angles $(\phi,\theta,\omega)$, while the integral over the function $s(\hat \mathbf{n})$ (the ``signal'') covers the whole sphere.  For band-limited functions \citet{WG2001} found a class of fast algorithms for this type of problem, computing the convolution over part or all of the sphere for some or all beam orientations.  These employ the spherical harmonic representation of the beam and signal, defined by 
\begin{equation}
  b(\hat \mathbf{n}) = \sum_{l=0}^{L} \sum_{m=-l}^{l} b_{lm} Y_{lm}(\hat \mathbf{n}),
\end{equation}
for the beam (similarly for the signal) where $L$ defines the band limit.
For the special case considering the whole sphere at all orientations, dubbed ``total convolution,'' the main result is
\begin{equation}
  T(\hat \mathbf{n},\omega) = \sum_{m_i = -L}^{L} T_{m_1 m_2 m_3} e^{m_1\phi + m_2 \theta + m_3 \omega} \label{eqn:T_FT}
\end{equation}
where
\begin{equation}
 T_{m_1 m_2 m_3} = \sum_{l=0}^{L} s_{lm_1} \Delta^l_{m_1 m_2} \Delta^l_{m_2 m_3} b^*_{l m_3}.
\end{equation}
Here $\Delta^l_{m_1 m_2} = d^l_{m_1 m_2}(\pi/2)$ are the reduced Wigner rotation matrices\footnote{See \citet[chapter 4]{Edmonds} for the definition of the $d$-functions and \citet{Risbo1996} for recursion relations for rapid computation.} evaluated at $\pi/2$.   For a uniform rectilinear grid in $(\phi,\theta,\omega)$,  equation (\ref{eqn:T_FT}) may be computed rapidly using a fast Fourier transform.  The main computational cost is in computing $T_{m_1 m_2 m_3}$, which scales as $O(L^4)$, compared to a computational cost of $O(L^5)$ for a brute force approach.  If the azimuthal content of the beam is limited, the scaling can be reduced by limiting $m_3$, yielding a scaling of $O(L^3 m_{3,{\rm max}})$.  In the applications in this work, the convolutions each require a few minutes to compute on modest computer hardware.  By conventional means, these computations would each take several days.

\subsection{Preparation of the harmonic representation}

Because the convolution may be computed efficiently in a band-limited harmonic representation, we need to construct a harmonic representation of the beam (and sky) which captures the relevant aspects of the function on the sphere, but does not introduce unphysical features, such as spikes at the sampling positions of the beam or ringing due to a sharp band-limit imposed in harmonic space.

Often the beam and sky maps are presented as equally spaced samples on isolatitudinal rings.  Compared to an arbitrary pixelization, this arrangement makes harmonic analysis more efficient: it allows the azimuthal portion of the transform to be computed by an FFT and reduces the number of Legendre function evaluations \citep[see e.g. HEALPix,\footnote{http://healpix.jpl.nasa.gov/}][]{Gorski2005}.  Additionally the samples may be distributed azimuthally on meridians.  
Because physical optics calculations of the telescope radiation pattern are resource-intensive, a practical examination of the sidelobes may sparsely sample the sphere.  Sidelobes typically subtend large solid angles, so this is not usually an issue, but it does have some implications for the evaluation of the beam harmonics.

The resuperposition of the harmonics (possibly on a different pixelization) will attempt to reproduce the original sampling of the beam as a collection of $\delta$-functions.  Due to the finite band limit, these $\delta$-functions are somewhat smeared.  However, if the sampling density of the output convolution anywhere exceeds that of the input grid, these sampling spikes will be visible.  This common problem arises when the sampling is sparse, or when the areal density of samples varies widely, such as on any grid on which the number of samples per latitude ring is constant, like the geographic grid.  Smoothing the harmonics can help address these issues with the beam representation, and yield a result closer to the physical beam.

A useful smoothing for grids  is an azimuth-direction-only smoothing:
\begin{equation}
  \bar b(\theta,\phi) = \int_0^{2\pi} \, d\phi' f(\phi-\phi') b(\theta,\phi)
\end{equation}
where the smoothing kernel has $\int_0^{2\pi} d\phi\, f(\phi) = 1$.
If we expand the kernel as $f(\phi) = \sum_m f_m \exp(-im \phi)$, the spherical harmonic coefficients of the smoothed function are,
\begin{eqnarray}
  \bar b_{lm} &=& \int d(\cos \theta)\, d\phi \,\bar b(\theta,\phi) Y^*_{lm}(\theta,\phi) \nonumber \\
&=& 2 \pi f_m b_{lm}.
\end{eqnarray}
This result is straightforward, and may be obtained by substitution of the definition of $\bar b$, expanding $f$ and $b$ in harmonic series, integrating over the azimuthal angles, using the orthogonality relation of associated Legendre polynomials, and summing over the resulting Kronecker $\delta$-functions.

One particularly useful example smoothing kernel is the unit-integral boxcar,
\begin{equation}
  f(\phi) = 
  \left\{
  \begin{array}{rl}
    (2\phi_0)^{-1}, & -\phi_0 \le \phi \le \phi_0 \le \pi; \\
    0, & \mbox{otherwise},
  \end{array} 
  \right.
\end{equation}
for which the harmonic coefficients are 
\begin{equation}
  2\pi f_m = 
  \left\{
  \begin{array}{rl}
    1, & m=0; \\
    \displaystyle{\frac{\sin(m\phi_0)}{m\phi_0}}, & m \ne 0.
  \end{array}
  \right.
\end{equation}

Another useful smoothing function is a symmetric Gaussian.  After smoothing with a symmetric Gaussian, the spherical harmonic coefficients are
\begin{equation}
\bar b_{lm} = f_l b_{lm}
\end{equation}
where 
\begin{equation}
f_l = \exp \left( - l (l+1) \sigma_{\rm smooth}^2/2 \right)
\end{equation}

Note that neither of these filters modify the beam monopole.  Thus the integral under the beam is unaffected by the smoothing.

\section{Applications} \label{sec:applications}

\subsection{Far-sidelobe pickup: \textit{CALISTO}}
For our first application of the above convolution method, we examine telescope far-sidelobe pickup.  In particular, we studied design concepts for a space-based infrared telescope. The Cryogenic Aperture Large Infrared Space Telescope Observatory, or \textit{CALISTO}, is tuned for the wavelength range 30--300 $\mu$m.  The whole optical system, including the $\sim 5$ m diameter primary mirror, is cooled to 4 K, and yields  a main beam resolution of $\sim 1.2''$--$12''$, depending on wavelength.  Based on the low emission of the cold optics, a very low system noise equivalent power can be achieved, $\sim 10^{-19}$ W Hz$^{-1/2}$ \citep{Goldsmith2006,Goldsmith2007}.

The purpose of this study was to calculate and compare the sidelobe pickup for three optical designs: (1) an on-axis design, where the secondary is held in front of the primary by three struts, (2) an off-axis Gregorian design, and (3) an off-axis Gregorian design with a 4 K cold stop placed between the detectors and the secondary.  The sidelobe pickup is a convolution of the sidelobes with the sky, and is a straightforward application of equation (\ref{eqn:convolution}).
The all-sky beams were calculated previously using GRASP9 in three latitudinal zones \citep{Goldsmith2007} (see table \ref{tab:beampix}), with decreasing angular resolution away from the main beam, which is centered at $\theta = 0^\circ$.   The on-axis beam contained $\sim 139$ million pixels, and the off-axis beams contained $\sim 193$ million pixels.

\begin{deluxetable}{lrrrr}
\tablewidth{\columnwidth} 
\tabletypesize{\small} 
\tablecaption{Beam parameters\label{tab:beampix}}
\tablecolumns{5}
\tablehead{& $\theta_{\rm max}$ ($^\circ$) & $\Delta \theta$ ($^\circ$) & $\Delta \phi$ ($^\circ$) & $\sigma_{\rm smooth}$ ($^\circ$)}

\startdata
  \\
on-axis &   1 & $5.0\times 10^{-4}$ & 0.0833 &  $5.0\times 10^{-4}$\\
 &   25 & $8.0\times 10^{-4}$ & 0.0833 & $1.0\times 10^{-3}$\\
 &   180 & $5.0\times 10^{-4}$ & 360.0 & 2.0\\ \\
off-axis &   3 & $5.0\times 10^{-4}$ & 0.0833 & $5.0\times 10^{-4}$\\
 &   40 & $1.0\times 10^{-3}$ & 0.0833 & $1.0\times 10^{-3}$\\
 &   180 & $1.4\times 10^{-3}$ & 5.0 & 2.0\\ 
\enddata
\tablecomments{Pixelization of the beam for on- and off-axis \textit{CALISTO} design cases from the GRASP computation.  Each is divided into three non-overlapping latitude zones, marked by the maximum colatitude ($\theta_{\rm max}$) in each zone.  The pixel centers in each zone are separated by $\Delta\theta$ and $\Delta \phi$.  In preparing the convolutions, the beams in each region were smoothed by an azimuthal boxcar filter of width $\Delta \phi$ and a symmetric Gaussian of width $\sigma_{\rm smooth}$.}
\end{deluxetable}

\begin{figure}
\includegraphics[height=0.33\columnwidth]{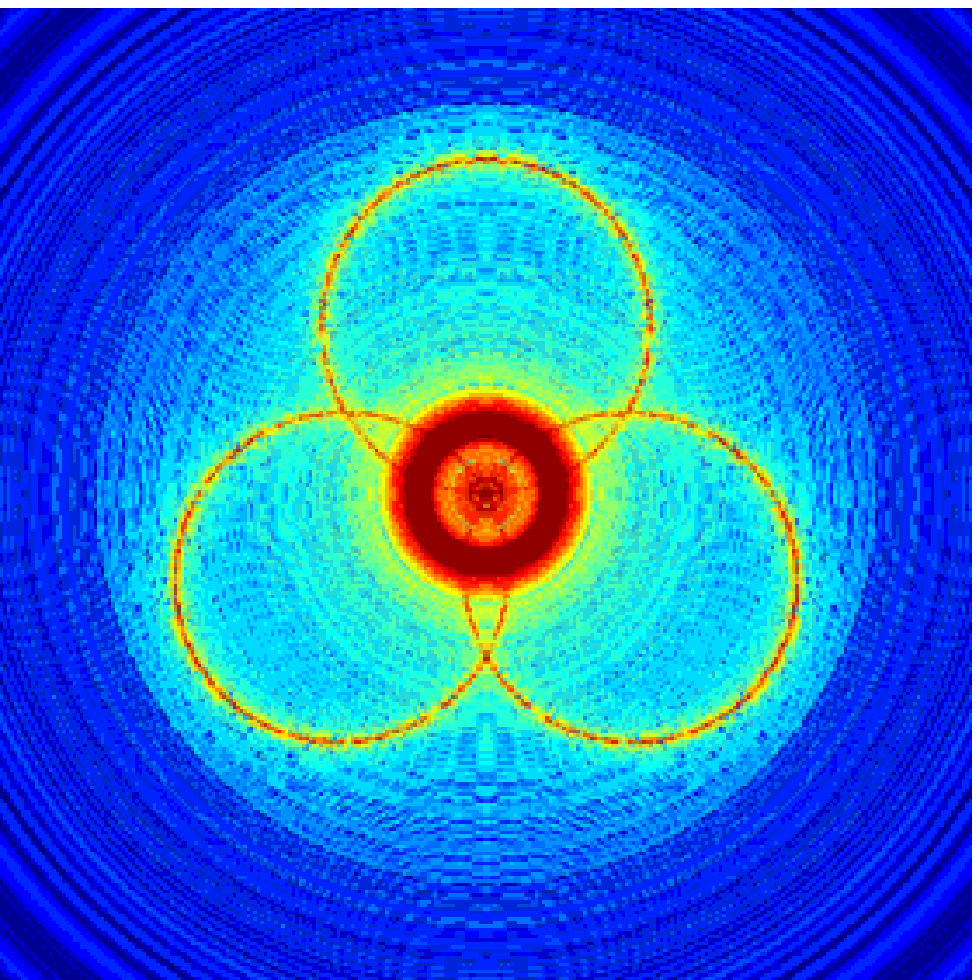}
\hfill
\includegraphics[height=0.33\columnwidth]{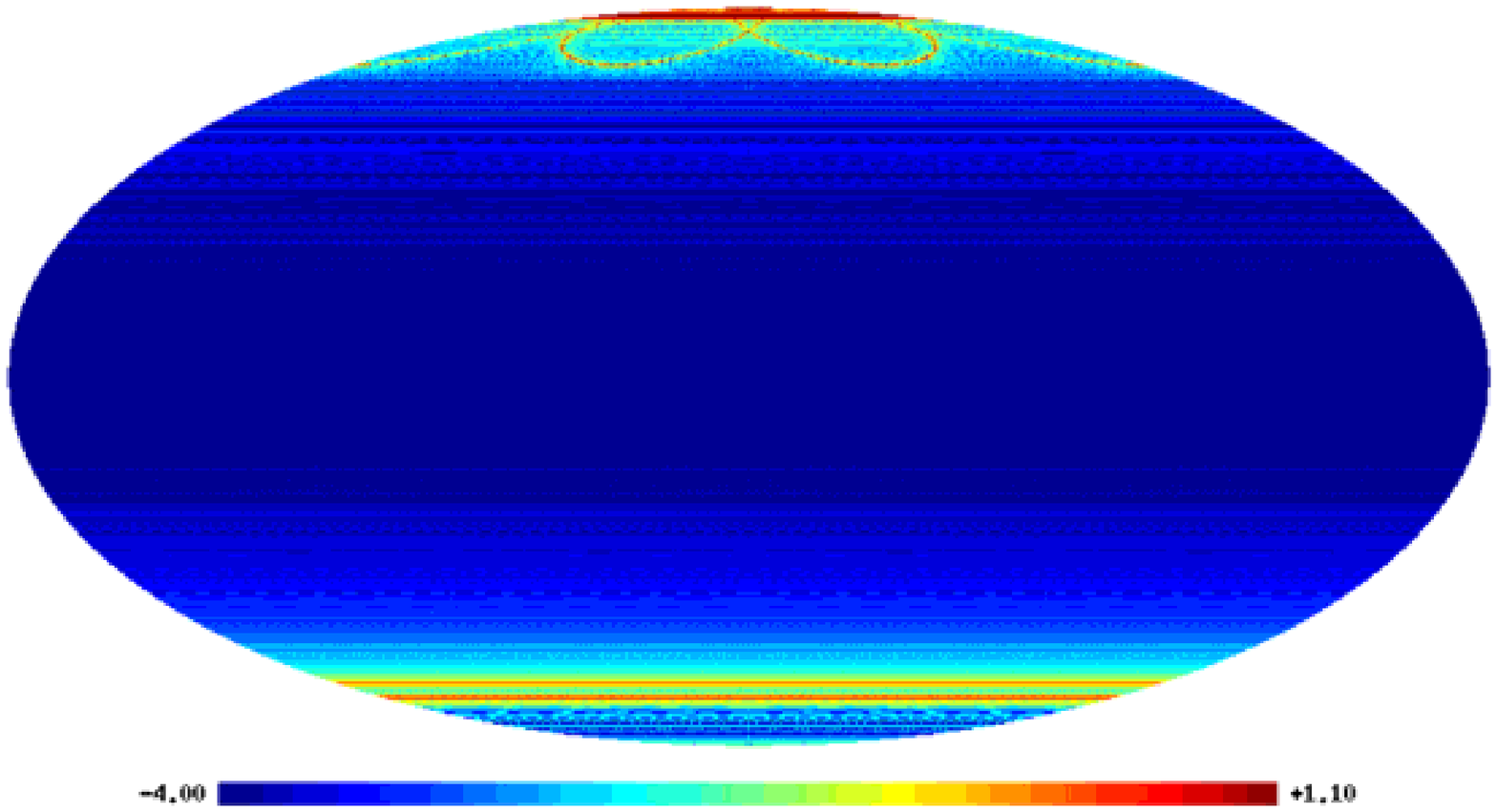}
\includegraphics[height=0.33\columnwidth]{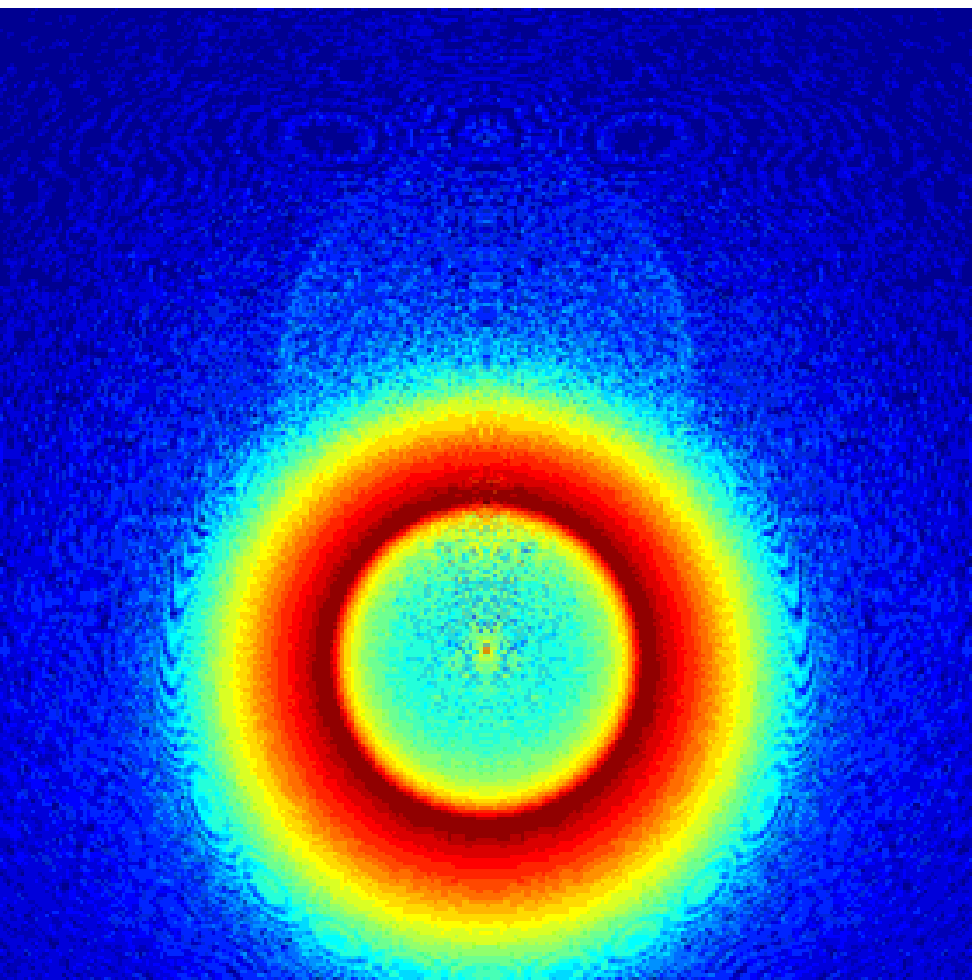}
\hfill
\includegraphics[height=0.33\columnwidth]{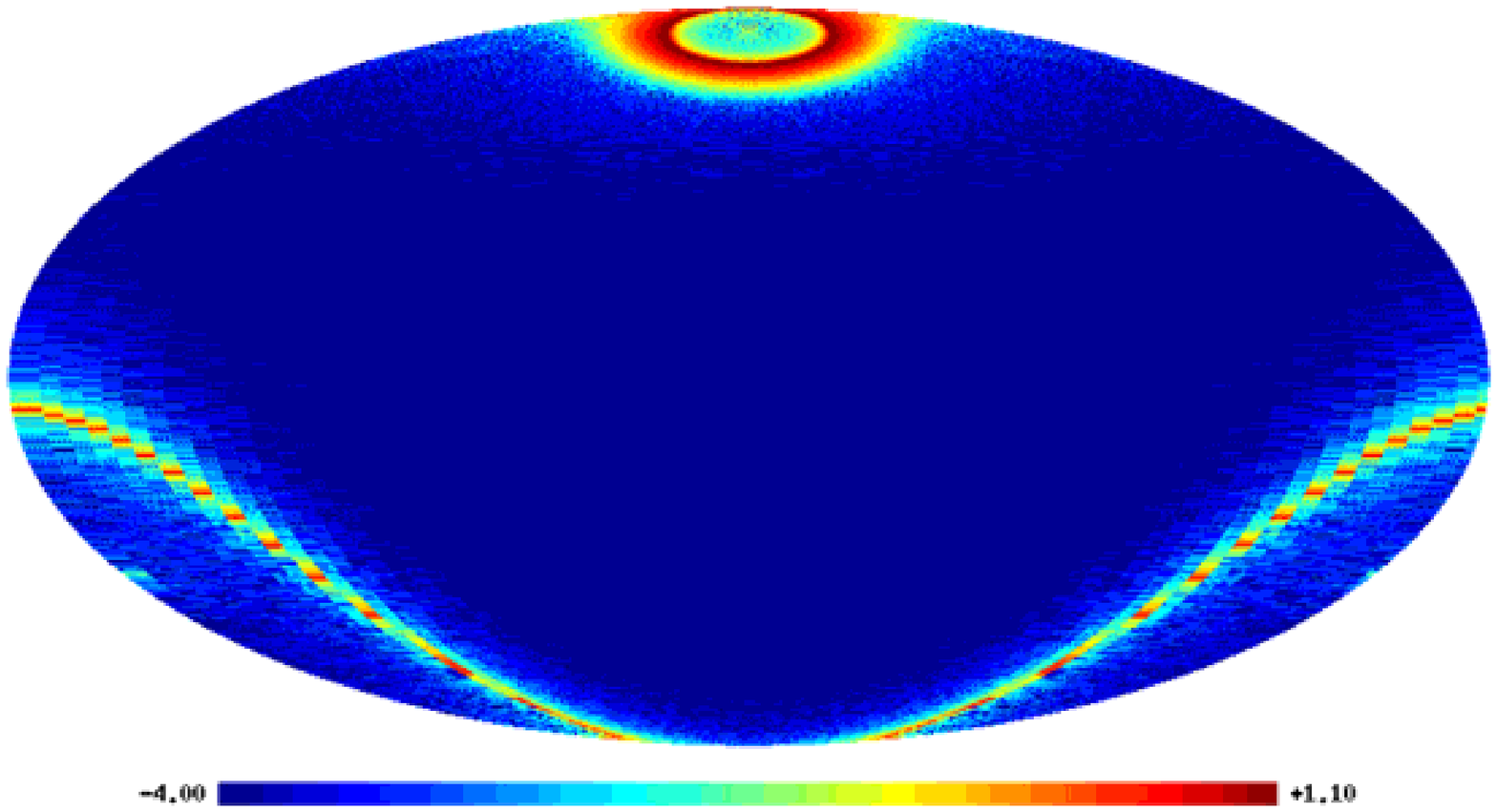}
\includegraphics[height=0.33\columnwidth]{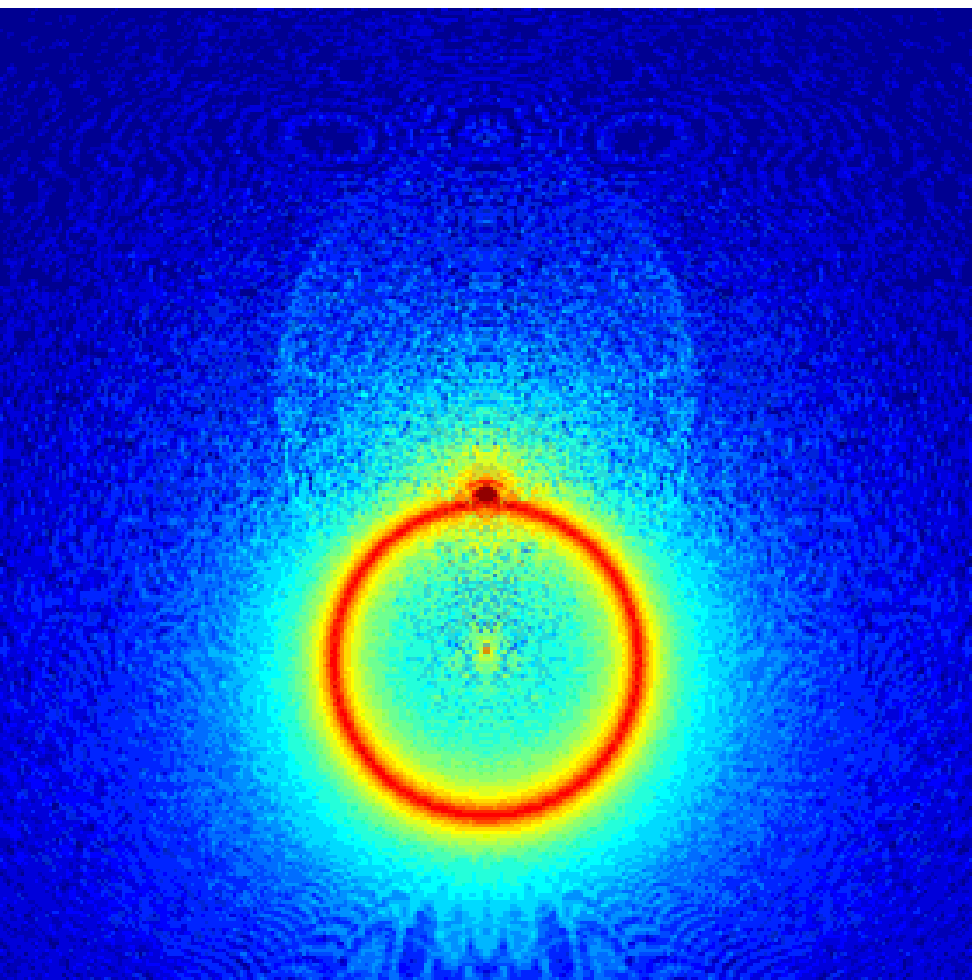}
\hfill
\includegraphics[height=0.33\columnwidth]{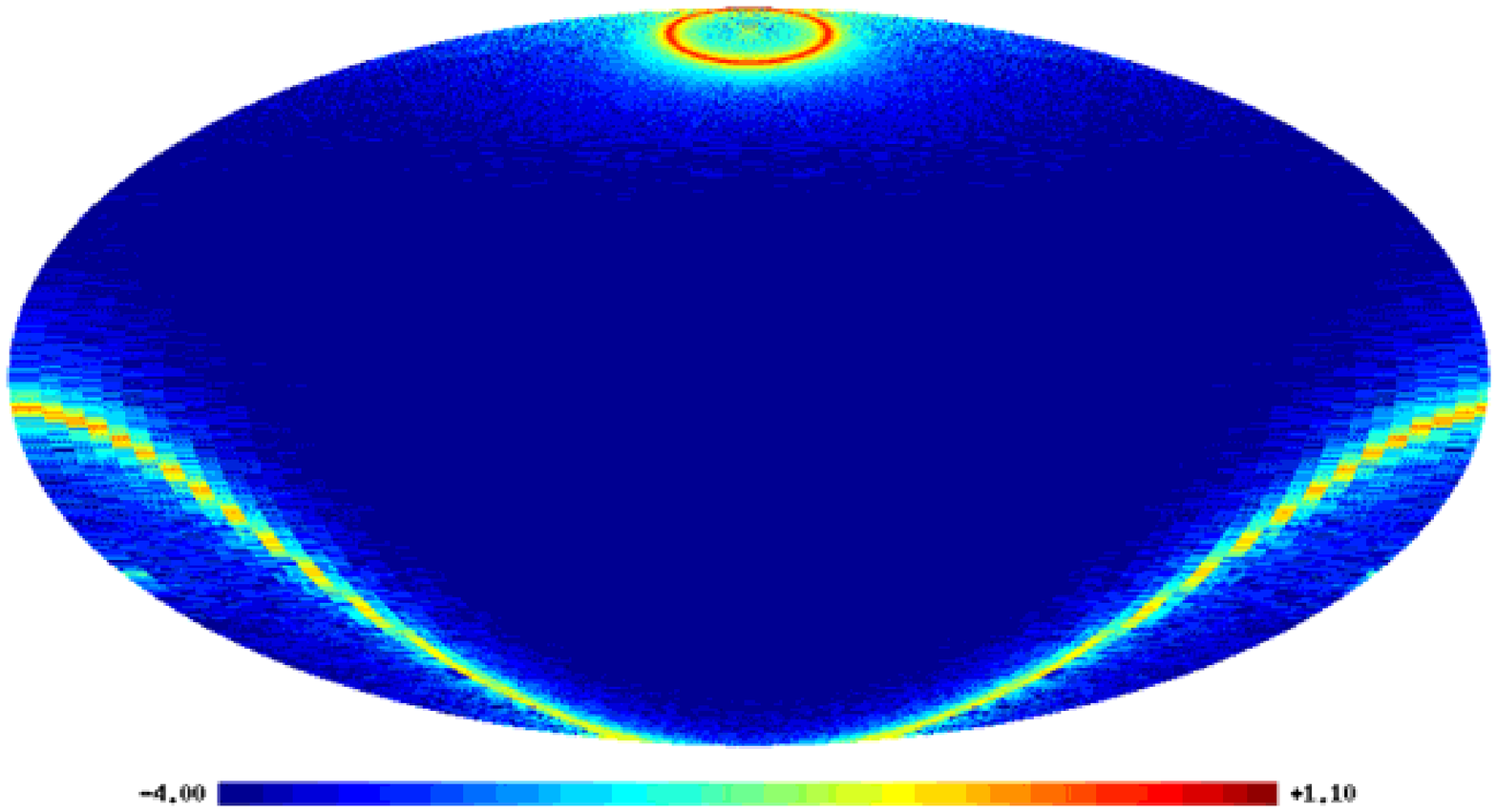}
\caption{The beams for three optical designs compared in the \textit{CALISTO} sidelobe study. To the left, the beams are displayed in the Gnomonic projection, with the main beam at the center.  To the right, the beams are shown in the Mollweide projection, with the main beam at the top.  Color scale indicates $\log_{10}(b)$ and saturates in the sidelobes of the on-axis design. (Top) The on-axis design, with notable reflections.  (Middle) The off-axis design with no cold stop, where the spillover from the secondary (closer to the main beam) and primary are not aligned along the optical axis.  (Bottom) The off-axis design with a cold stop is similar, with reduced sidelobes.}
\label{fig:beampics}
\end{figure}

In figure \ref{fig:beampics}, we show the beams from each of the three designs.  Each design shows spillover from the primary mirror (where the spillover is opposite the axis of the main beam) and the secondary (where it is closer to the main beam).  In the on-axis design, the support struts cause a 3-fold symmetric reflection pattern.  The off-axis designs have lower sidelobes, particularly in the case including a cold stop.

To suppress the main beam, in each case we multiplied the inner 20 arcminutes by an azimuthally symmetric Gaussian weight function, smoothly rising to unity at $\theta_0 = 20'$ with a standard deviation of $\sigma_\theta = 2'$:
\begin{equation}
w(\theta) = \left\{
\begin{array}{ll}
\exp\left(-(\theta - \theta_0)^2/2\sigma_\theta^2 \right) & \theta < \theta_0 \\
1 & \theta > \theta_0
\end{array} \right.
\end{equation}

The harmonic components were computed separately in the three latitude zones.  We wrote a custom code to compute the spherical harmonic transforms, using a harmonic transform library developed within the Planck collaboration \citep{Reinecke2006personal} and MPI\footnote{http://www-unix.mcs.anl.gov/mpi/} for multiprocessor computation.  We experimented with tapering the transition between the latitudinal zones for spherical harmonic transformation, but found this was not required to capture the large scale features of the beam.

\begin{figure*}
\includegraphics[width=0.49\textwidth]{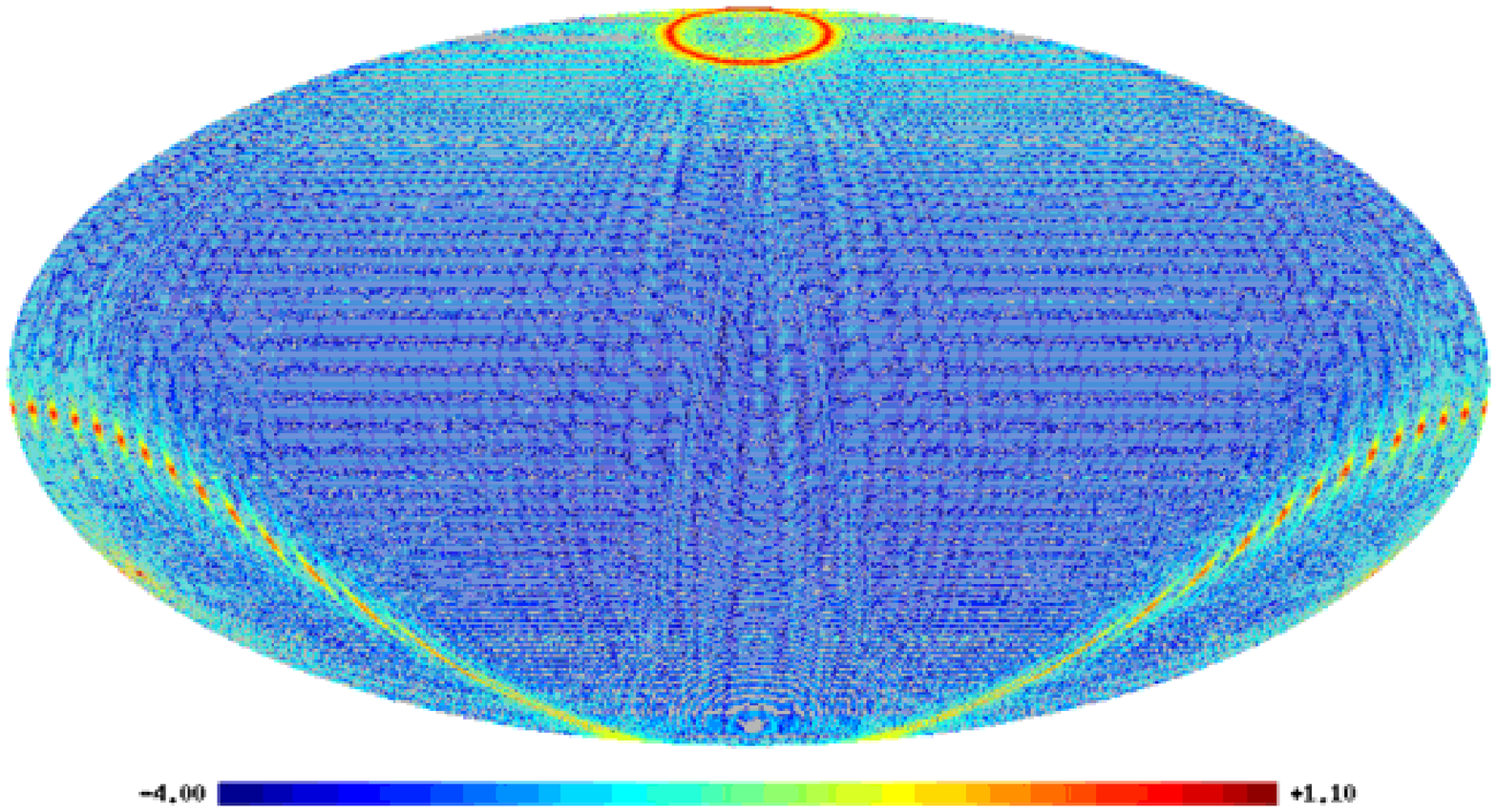} \hfill
\includegraphics[width=0.49\textwidth]{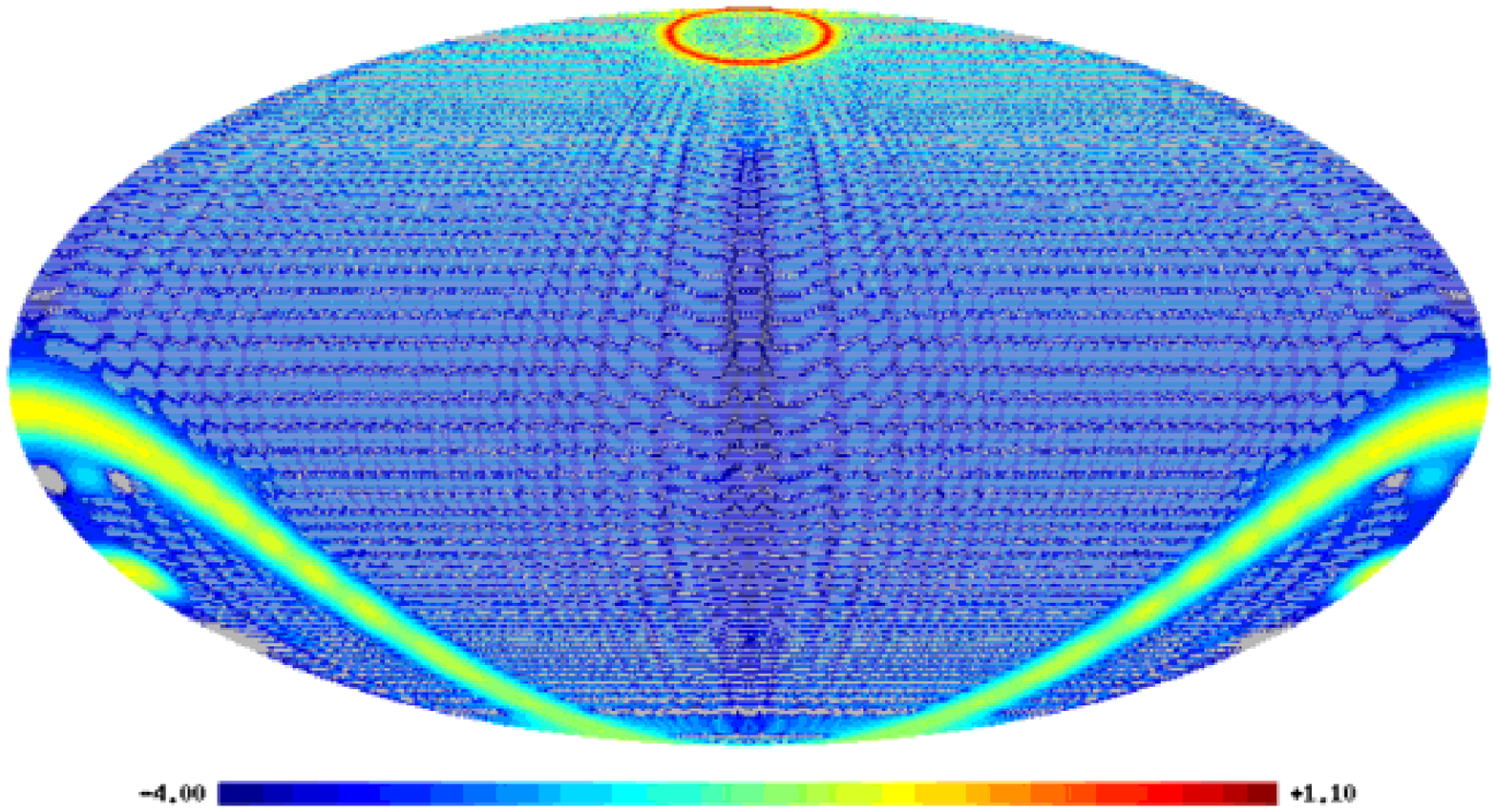}
\caption{Resuperposition of harmonic components for the off-axis, cold stop case.  Color indicates $\log_{10}(b)$.  Values less than zero are shown in gray, and are an unphysical consequence of the finite band limit, but are small enough to have little consequence in the convolution.  (Left) Without smoothing; note the shape of the low-resolution sidelobe, like a $\delta$-function comb.  (Right) The smoothed representation. }
\label{fig:resup}
\end{figure*}

In the spherical harmonic basis, the three zones were then separately smoothed with a pixel-scale azimuthal boxcar and a symmetric Gaussian, and then summed. (See table \ref{tab:beampix} for smoothing parameters.)  The smoothing is necessary to suppress spikes in the resuperposition of the beam,  caused simply by the input  pixelization  (from the GRASP computation).  Figure \ref{fig:resup} show the result of this smoothing.  Without it, the spikes around the centers of the original pixels propagate into the convolution, leading to unphysical structure in the result.

\begin{figure*}
\centerline{\includegraphics[width=0.49\textwidth]{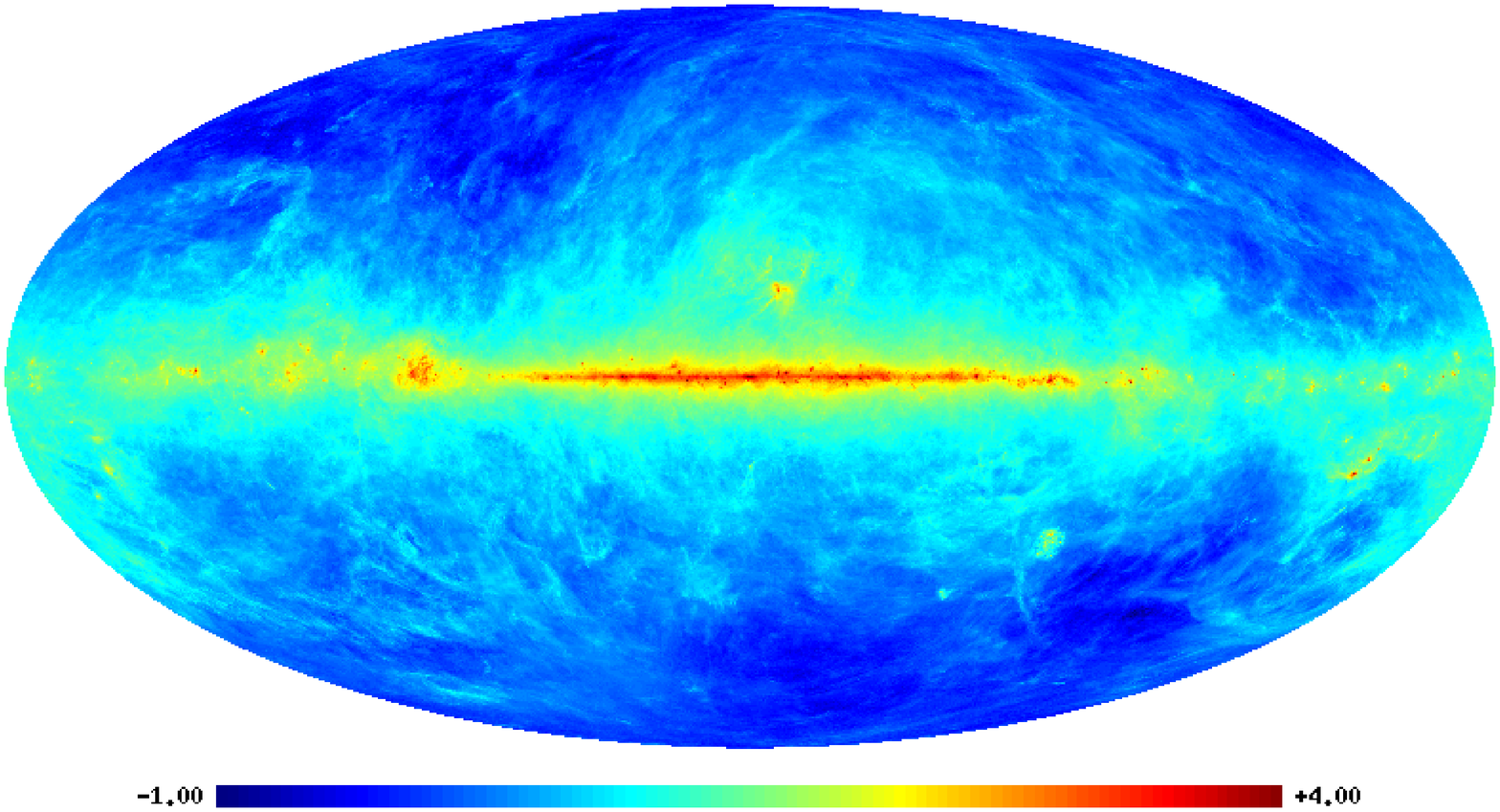}}
\includegraphics[width=0.49\textwidth]{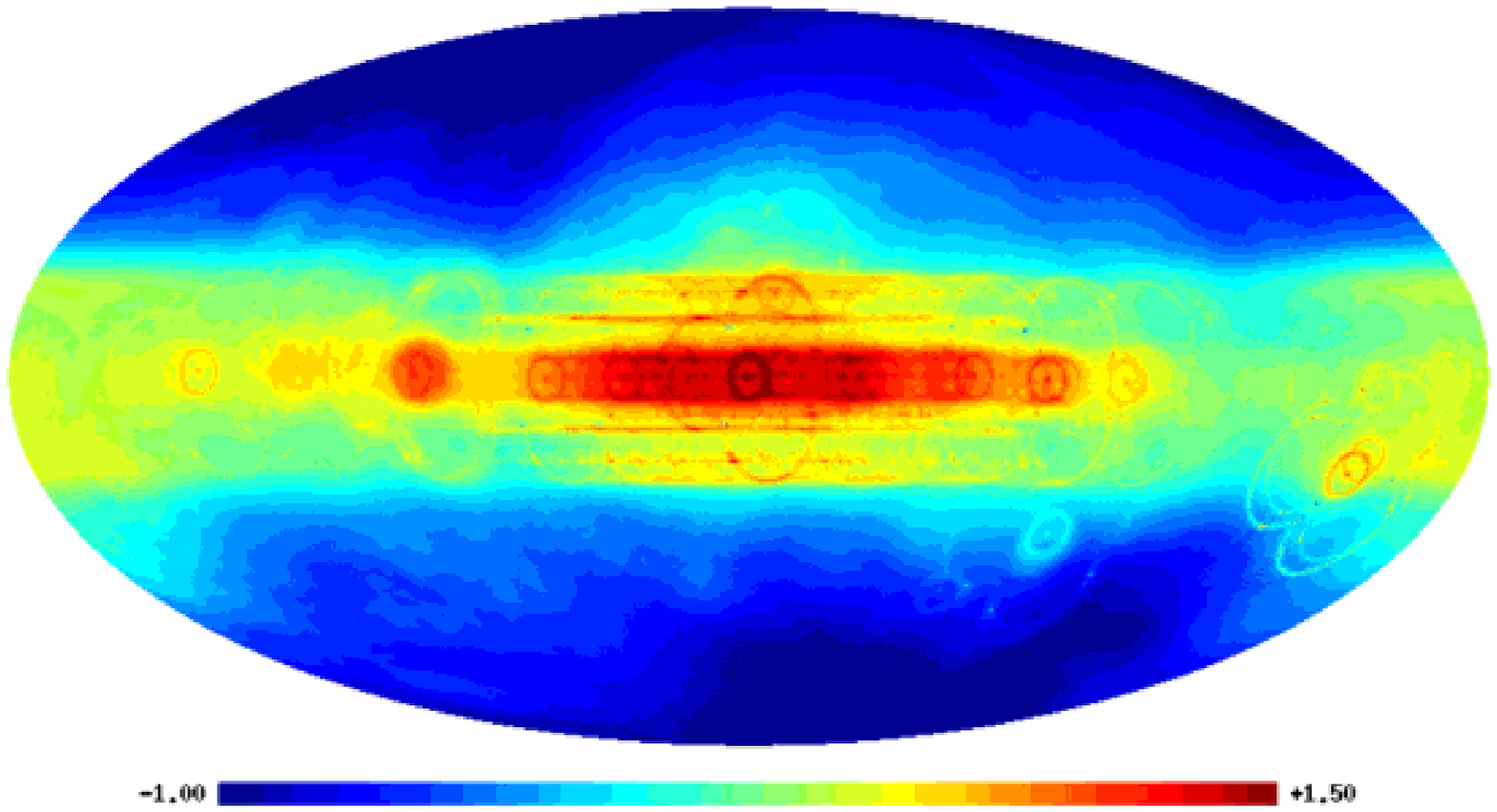} \hfill
\includegraphics[width=0.49\textwidth]{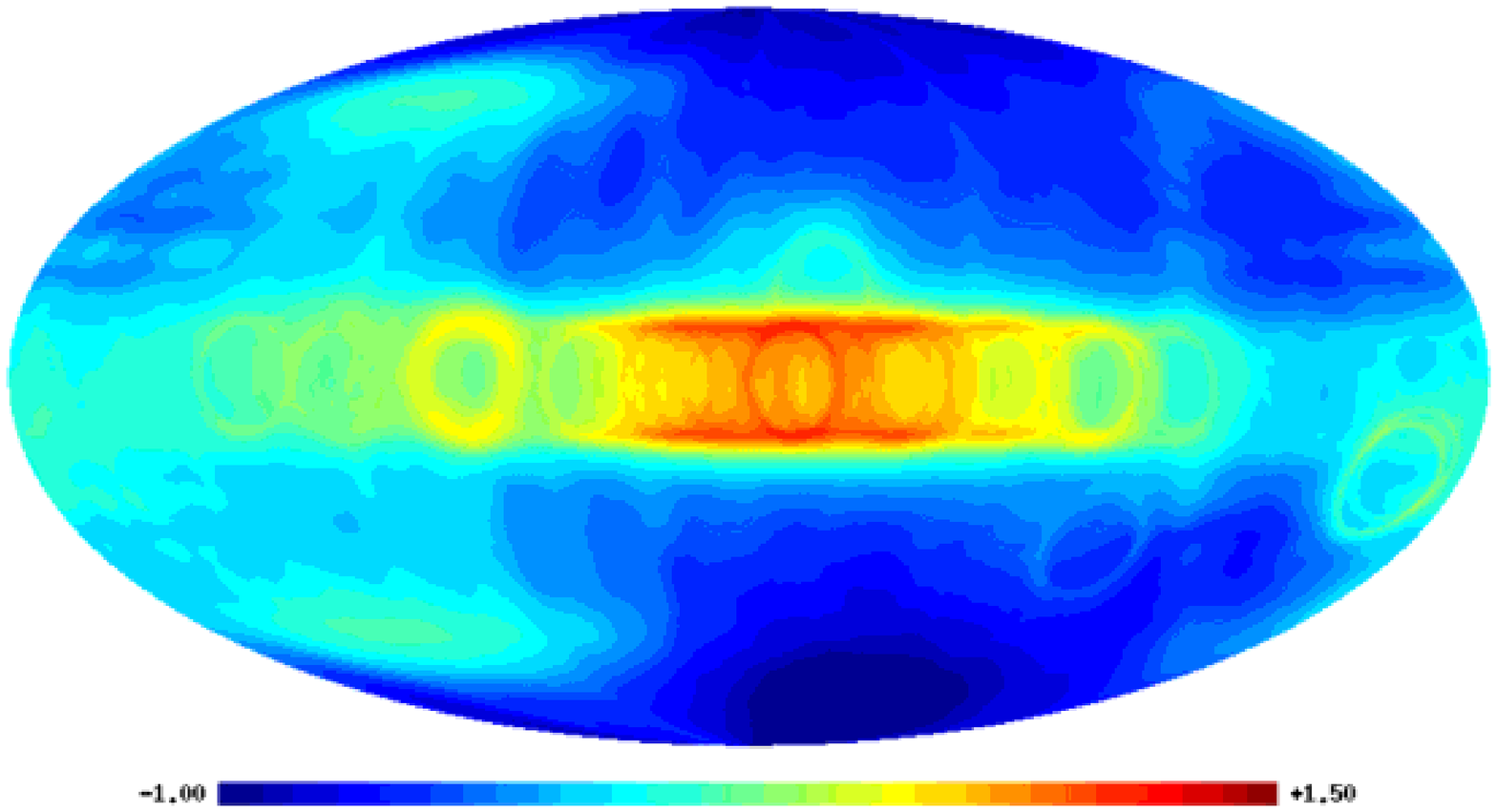}
\includegraphics[width=0.49\textwidth]{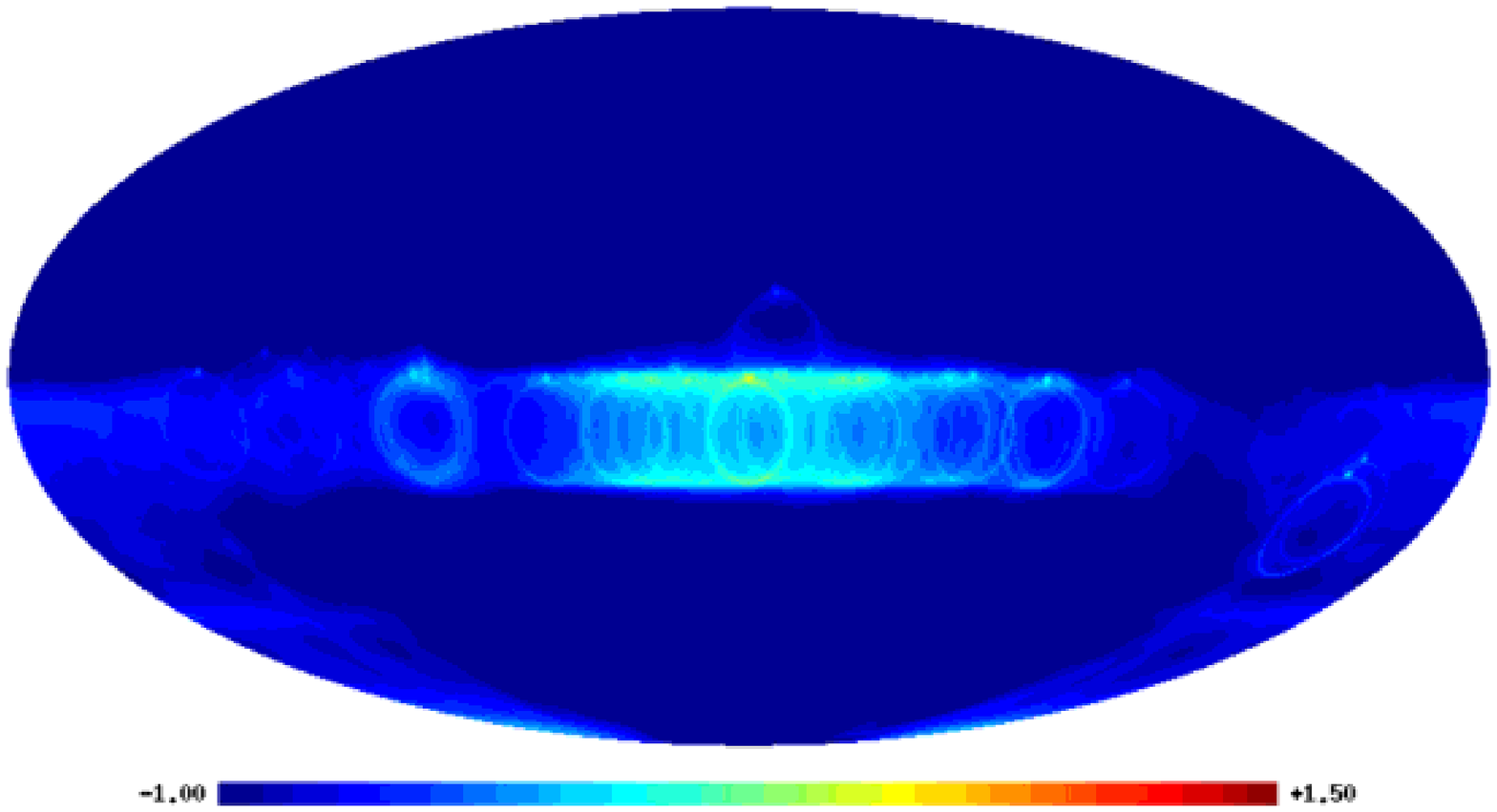} \hfill
\includegraphics[width=0.49\textwidth]{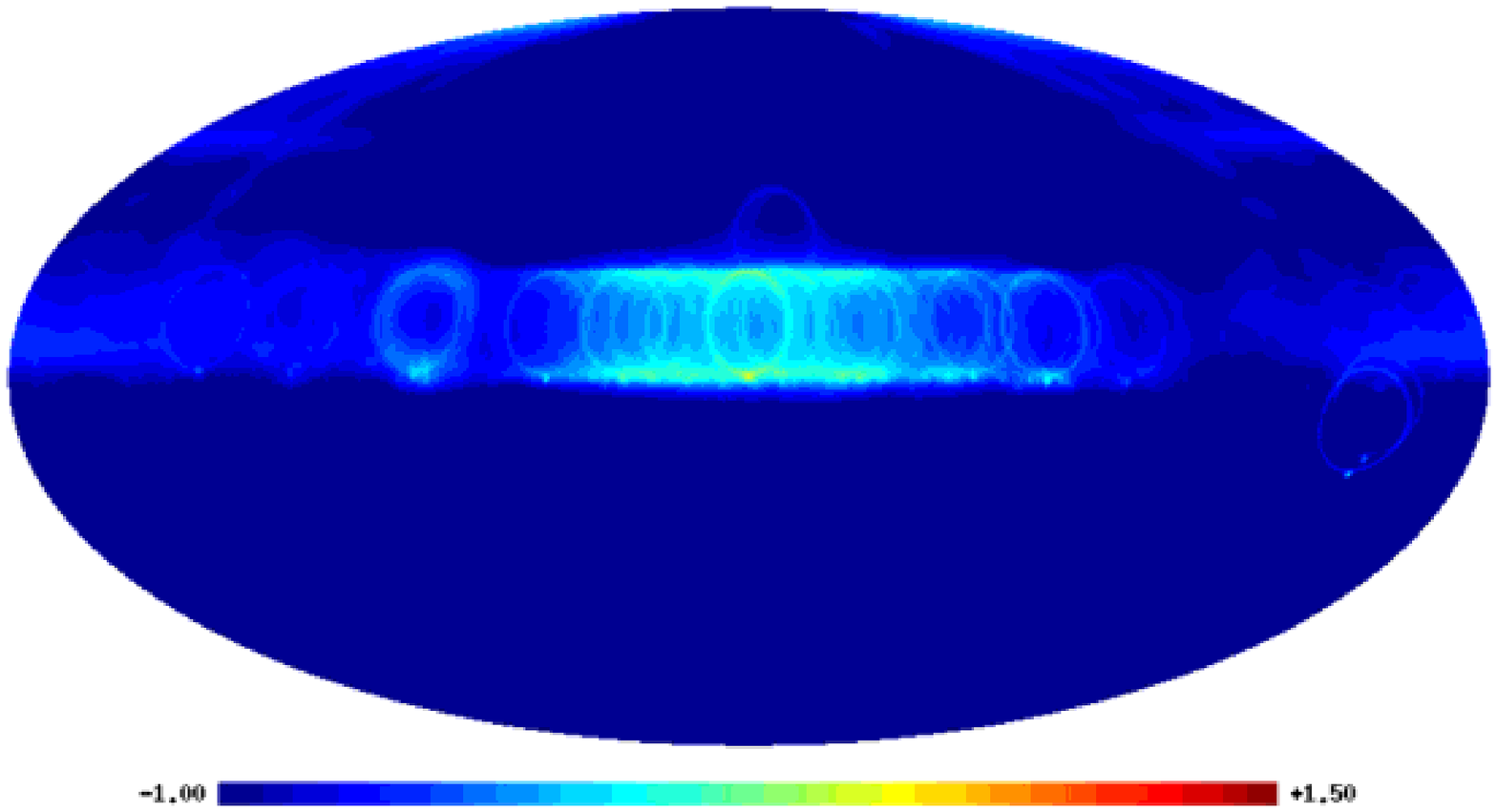}
\caption{The sky and the sidelobe pickup for prospective \textit{CALISTO} designs. Color indicates $\log_{10}( \mbox{MJy/sr} )$.  For the convolutions, slices through the convolved sky at fixed orientations are depicted as spheres in the Mollweide projection.  (Top) \citet{SFD1998} galaxy model at 100 microns.  (Middle-left) A slice through the convolution for the on-axis design.  (Middle-right) A slice for the off-axis design with no cold stop.  (Bottom-left) The off-axis design with a cold stop, with the spacecraft oriented to minimize sidelobe contamination in the northern galactic hemisphere.  (Bottom right) The same, but oriented to minimization contamination in the southern hemisphere.}
\label{fig:pickup}
\end{figure*}

To perform the convolution, we use the \citet{WG2001} ``totalconvolver'' software, in its implementation optimized by \citet[section 6.5]{LevelS}.  We limit the harmonic content to $L \leq 512$, although we limit the azimuthal beam index $m_3 \leq 256$.  On a 4-processor, 2.4 GHz Opteron machine, the convolution completes in 3 minutes.  We repixelize the convolution into a set of $N_{\rm side} = 512$ ($7'$ pixel) HEALPix maps, one for each azimuthal $(\omega)$ rotation of the beam.

For visualization, we take 2-dimensional slices from  the 3-dimensional convolution in $(\phi,\theta,\omega)$.  At fixed $\omega$, we plot the convolution as a map on the surface of the sphere defined by $(\phi,\theta)$.  That is, each map pixel displays the \textit{entire} contribution to the instrument response from the beam sidelobes, when the beam center is directed at the pixel and rotated about that axis according to $\omega$.  For a $\delta$-function signal on the $\theta = \pi/2$ plane, maps of these slices look like images of the beam, pinwheeling as the orientation angle is incremented.   This can be misleading, causing the viewer to think the map is a composite image of beams instead of a slice of the convolution.  This distinction becomes clearer near the pole, where the coordinate singularity of this representation distorts the beam image.  For example, the convolution of any beam with a $\delta$-function at the north pole will appear azimuthally symmetric in these slices with fixed $\omega$.

In Figure \ref{fig:pickup}, we show sample convolutions with the 100 micron dust map of \citet{SFD1998}.  The on-axis case shows the most sidelobe pickup.  More importantly, rotating the instrument about the optical axis has little impact on the structure of the sidelobe pickup.  The off-axis cases, however, give much better control over the placement of the sidelobes.  Because the sidelobes are not symmetric about the optical axis, there is the opportunity to orient them away from the galaxy.  For most of the sky with galactic latitude $|b| > 25^\circ$, the cold-stop design allows an orientation which limits the foreground contribution to $ < 0.1 $ MJy sr$^{-1}$, or roughly a few hundredths of a mJy per main beam.

\subsection{Scanning with asymmetric main beams: \textit{Planck}} \label{sec:asym_beam}
For a scanning experiment with asymmetric beams, the instrument response for every observation can be challenging to compute.  The resulting maps depend both on the beam shape, and the details of the scanning path and orientation, which can serve to symmetrize the effective beam.
Because the fast convolution method simultaneously computes the beam response in all possible configurations, it provides an attractive approach.  Here the challenge is that for a small main beam, the band limit must be high to fully capture the beam structure.  However, since the main beam often has limited azimuthal content (see \citet{Fosalba2002} for the case of an elliptical Gaussian), even with a high band limit, the technique of section \ref{sec:method} may still be tractable.  This approach is routinely used in the Planck simulation pipeline \citep{LevelS} to incorporate the effects of asymmetric beams.  

Here we quantify the impact of beam asymmetries on one of the main observables for \textit{Planck}, the temperature power spectrum.  We use simulated data from the 30 GHz channel of the Low Frequency Instrument (LFI), with 4 detectors at 30 GHz \citep[the simulations are detailed in][]{Trieste}.  These beams have a full-width half maximum of $\sim 33'$, and have substantial asymmetry, since the horns are near the edge of the focal plane \citep[the \textit{Planck} ``Bluebook,''][]{PlanckBluebook}.  The beams are modeled as elliptical Gaussians; as an indication of realistic beam asymmetry, we take an axis ratio similar to the K band from \textit{WMAP} \citep{2003ApJS..148...39P}.  We compare to axisymmetric Gaussian beams with the same effective area.
 
Noiseless time-ordered data were generated with the convolutions of the beams and the CMB sky, interpolating the response of each detector along the 1-dimensional scan path through the 3-dimensional convolution.  The realistic scan pattern had the satellite spin axis pointing along a cycloidal path around the anti-sun direction for one year.
The convolution band limit was $L = 3000$ and the maximum azimuthal index of the beam was $m_{3,{\rm max}} = 14$.
This procedure is handled by software in the Planck simulation pipeline.  Here, the time-ordered data are simply binned to make a pixelized map of the sky (HEALPix $N_{\rm side} = 512$), then the power spectrum is evaluated.

\begin{figure}
\includegraphics[width=\columnwidth]{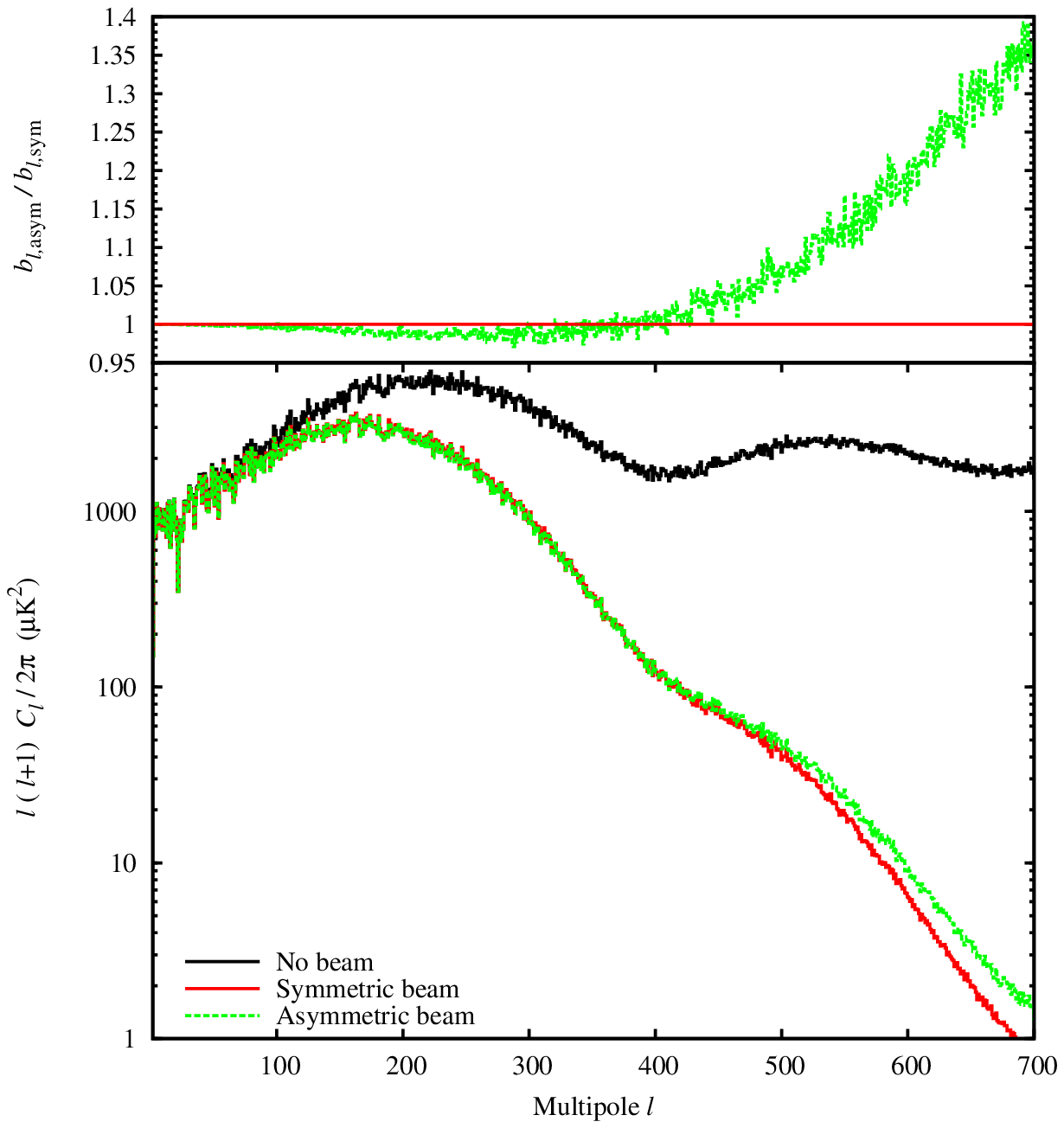}
\caption{The impact on the power spectrum due to asymmetric beams.  (Bottom) the power spectrum of the CMB, with no beam applied, 
 after simulated scanning with the four detectors of the Planck LFI 30 GHz channel with symmetric Gaussian beams with FWHMs of $\sim 32'$, and scanned with asymmetric elliptical Gaussian beams with the same effective area.
(Top) the ratio of the asymmetric to symmetric beam transfer functions, corresponding to the square root of the power spectrum ratio.}
\label{fig:cmbpowerspectrum}
\end{figure}

Figure \ref{fig:cmbpowerspectrum} compares power spectra.  The asymmetric beams are compared to symmetric Gaussian beams scanned the same way.  At very large scales, the beam and its shape are immaterial.  At intermediate scales, up to $l \sim 400$, the asymmetric beam suppresses power more than the symmetric one.  By contrast, at small scales, the symmetric beam suppresses more power, since the short axis of the asymmetric beam is able to sample smaller scales.

\subsection{Orientable filters}

For our third application, we examine fast convolution as a method to search for specified patterns in data on the sphere. In several works which process the sky with steerable wavelets \citep{
2006MNRAS.368..226B,
2006MNRAS.369...57C,
2006ApJ...636L...1L,
2006MNRAS.369.1858M,
2007arXiv0704.3158M,
2007arXiv0704.0626M,
2006MNRAS.365..891V,
2006NewAR..50..880V,
2007arXiv0704.3736V,
2006PhRvL..96o1303W,
2007arXiv0704.3144W,
2007arXiv0706.2346W},
the \cite{WG2001} convolution mechanism underlies the orientable filtering, making these methods computationally efficient and numerically tractable. The spirit of the formalism here is similar to \citet{2006ApJ...643..616J}.  

Suppose our data set is a scalar function $d(\hat \mathbf n)$ on the sphere, the sum of the signal and Gaussian noise with noise covariance $C(\hat \mathbf n,\hat \mathbf n')$.   For the case that we know the shape of a signal template, $s_0(\hat \mathbf n)$, and we want to estimate the amplitude ($A$) of this template, our model for the signal is $ A [D s_0]( \hat \mathbf n')$, where $D = D(\hat \mathbf{n},\omega)$ is the rotation operator.  For computational ease, we briefly switch to a vector notation, viewing our data as $d(\hat \mathbf n) = \mathbf{d}$. The likelihood is given by
\begin{eqnarray}
- 2 \log \mathcal{L}(A|\mathbf{ d},\hat \mathbf{n},\omega) &=& (\mathbf d - A[D \mathbf s_0])^\dag \mathbf C^{-1} \nonumber \\
& & (\mathbf d - A[D \mathbf s_0])
\end{eqnarray}
plus an immaterial constant.  We can write the maximum likelihood estimate of $A$:
\begin{equation}
  \tilde A(\mathbf{ d},\hat \mathbf{n},\omega) = \frac{[D \mathbf s_0]^\dag  \mathbf C^{-1} \mathbf{ d}}{[D \mathbf s_0]^\dag   \mathbf C^{-1} [D \mathbf s_0] } \label{eqn:match}
\end{equation}
Viewed as a filter acting on $\mathbf d$ to produce $A$, this is known as a matched filter.

Switching back from our vector notation, the numerator is easily written as a convolution (equation \ref{eqn:convolution}):
\begin{eqnarray}
\int d\hat \mathbf n'd\hat \mathbf n''\  [D  s_0]^*\!(\hat \mathbf n') C^{-1}(\hat \mathbf n', \hat \mathbf n'') d(\hat \mathbf n'') \nonumber \\ 
= \int d\hat \mathbf n' \  [D(\hat \mathbf{n},\omega)  s_0]^*\!(\hat \mathbf n') (C^{-1}d)(\hat \mathbf n').   \label{eqn:match_numerator}
\end{eqnarray}
If the noise is statistically homogeneous and isotropic, the integration over the inverse covariance is most easily done in spherical harmonic space,
\begin{equation}
  (\mathbf C^{-1}\mathbf d)_{lm} = \frac{d_{lm}}{C_l}.
\end{equation}
Equation (\ref{eqn:match_numerator}) may now be computed using the total convolution method.

The denominator  of equation \ref{eqn:match} is 
\begin{equation}
\int d\hat \mathbf n'd\hat \mathbf n''\  [D  s_0]^*\!(\hat \mathbf n') C^{-1}(\hat \mathbf n', \hat \mathbf n'') [D  s_0](\hat \mathbf n'')
\end{equation}
Again if the noise is isotropic, we may drop the common $D$ rotations of the signal template and evaluate the denominator in harmonic space, where it is equal to
\begin{equation}
  \sum_{l=0}^{L} \sum_{m=-l}^{l}  \frac{{s_{0,lm}}^* s_{0,lm}}{C_l} =   \sum_{l=0}^{L} (2l+1) \frac{C_l^{s_0}}{C_l}.
\end{equation}

\begin{figure*}
\begin{center}
\hfill \includegraphics[width=0.25\textwidth]{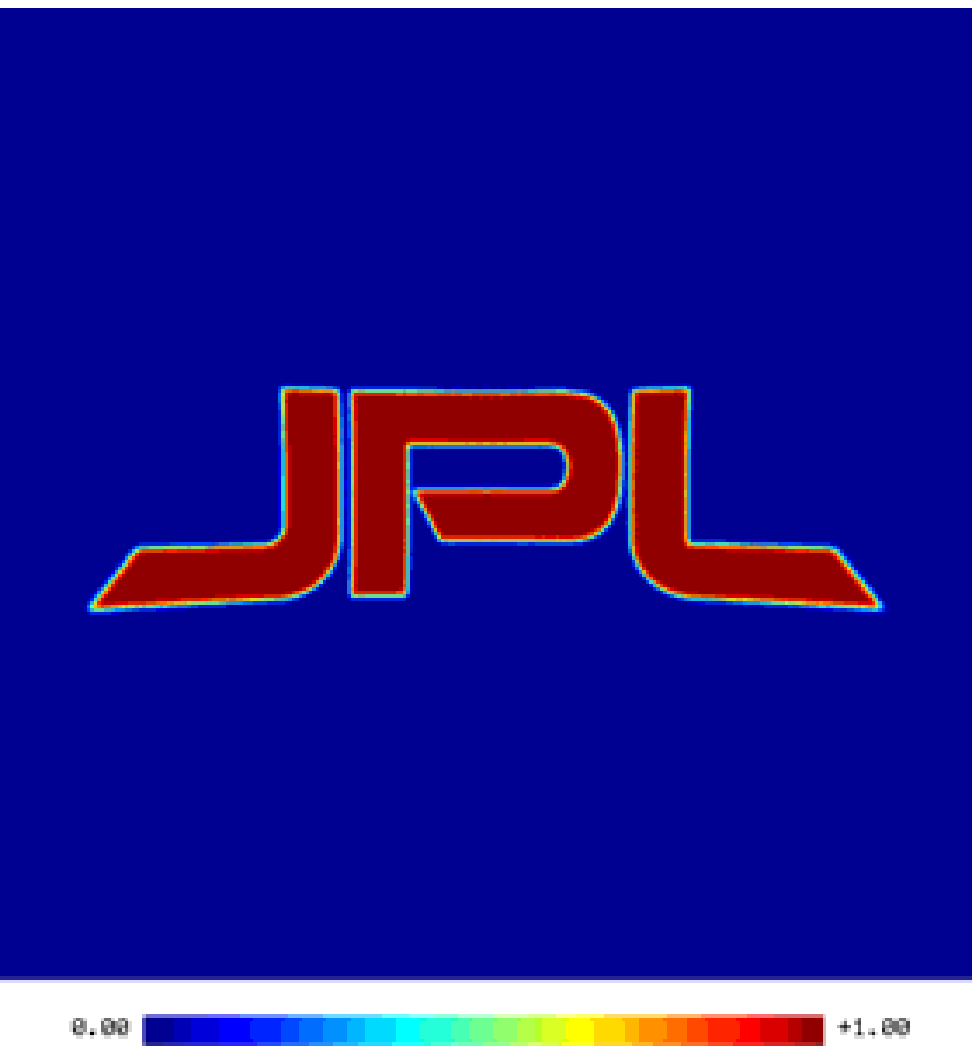} \hfill
\includegraphics[width=0.49\textwidth]{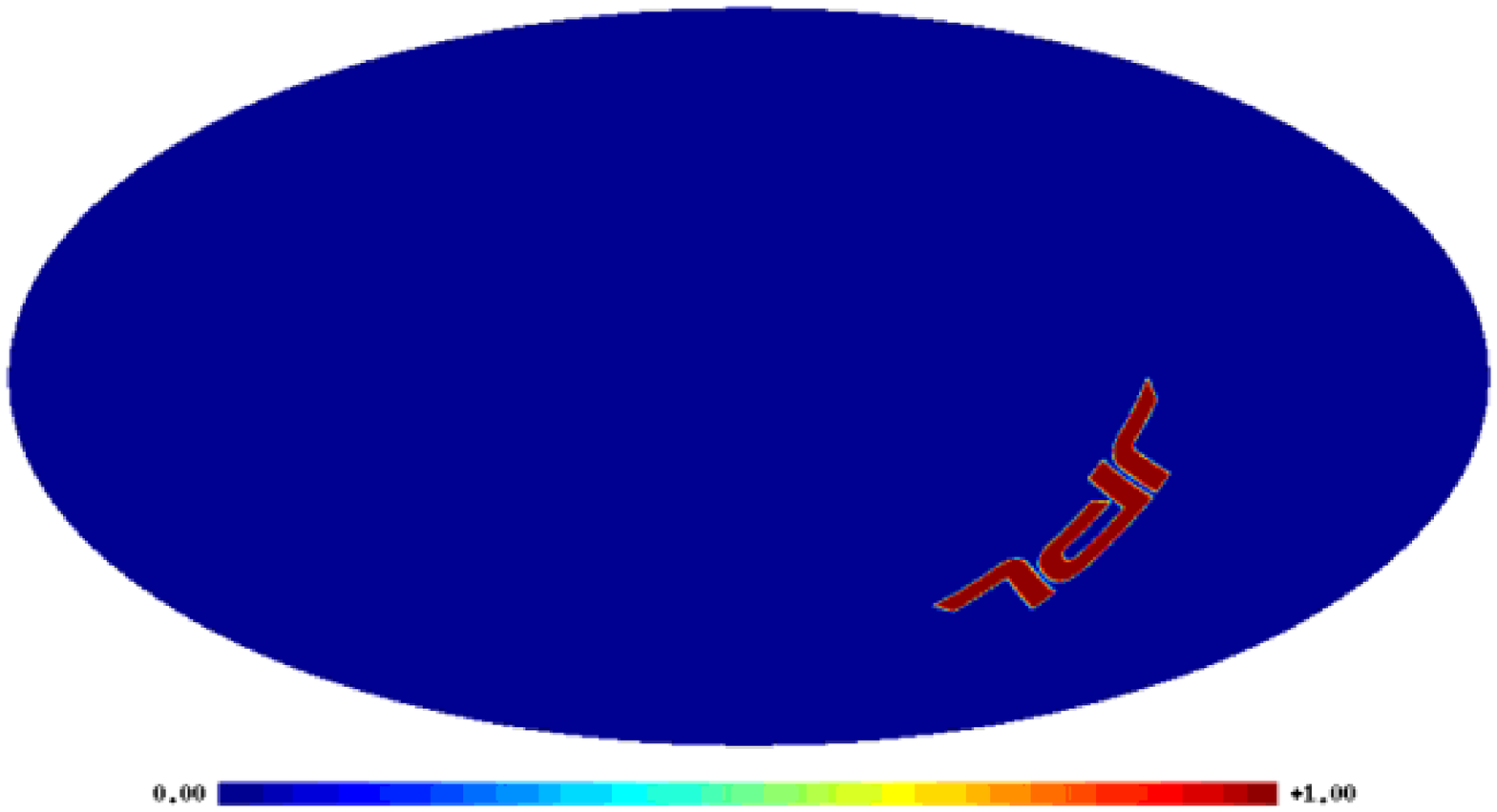}
\includegraphics[width=0.49\textwidth]{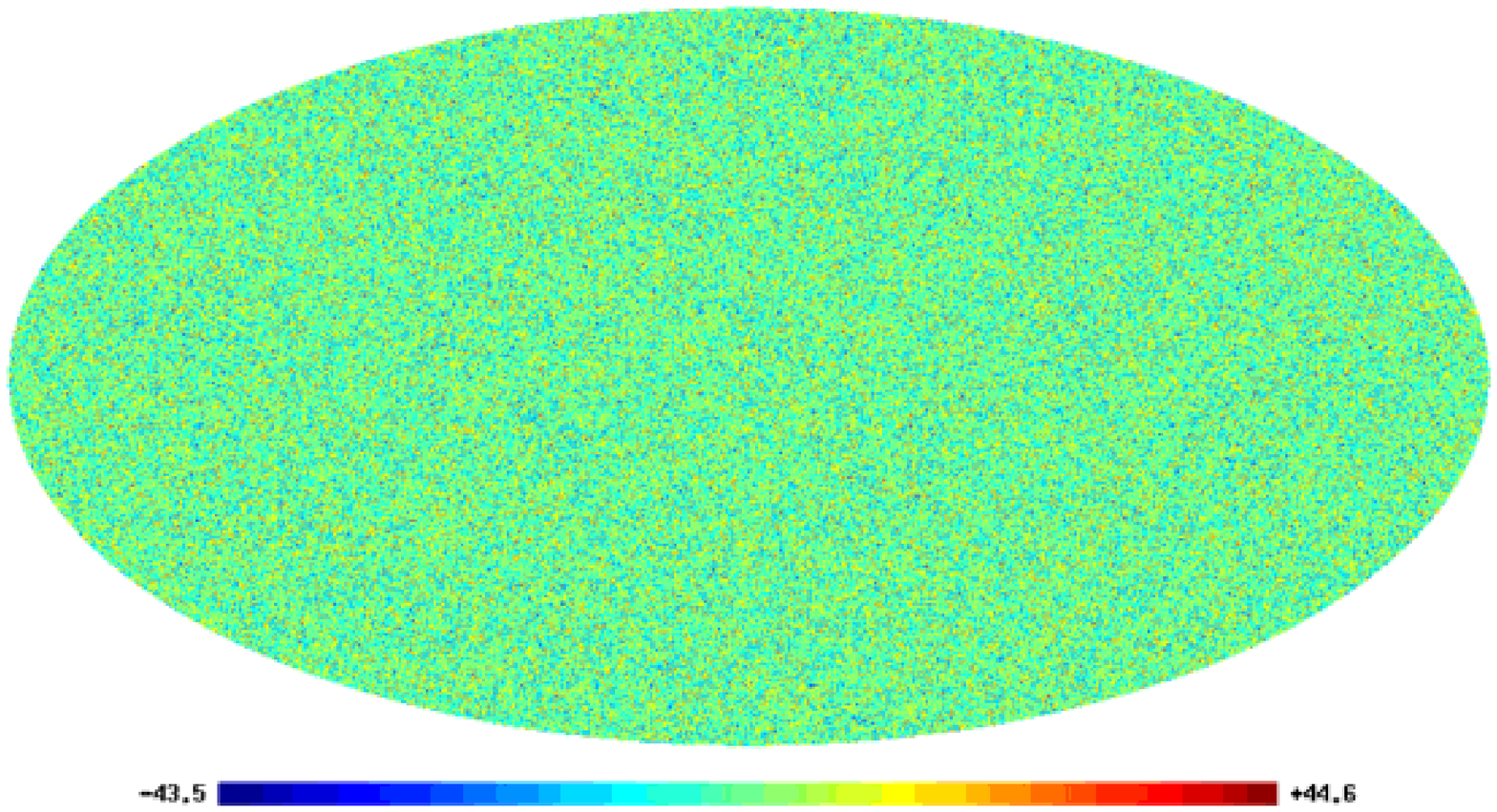}  \hfill
\includegraphics[width=0.49\textwidth]{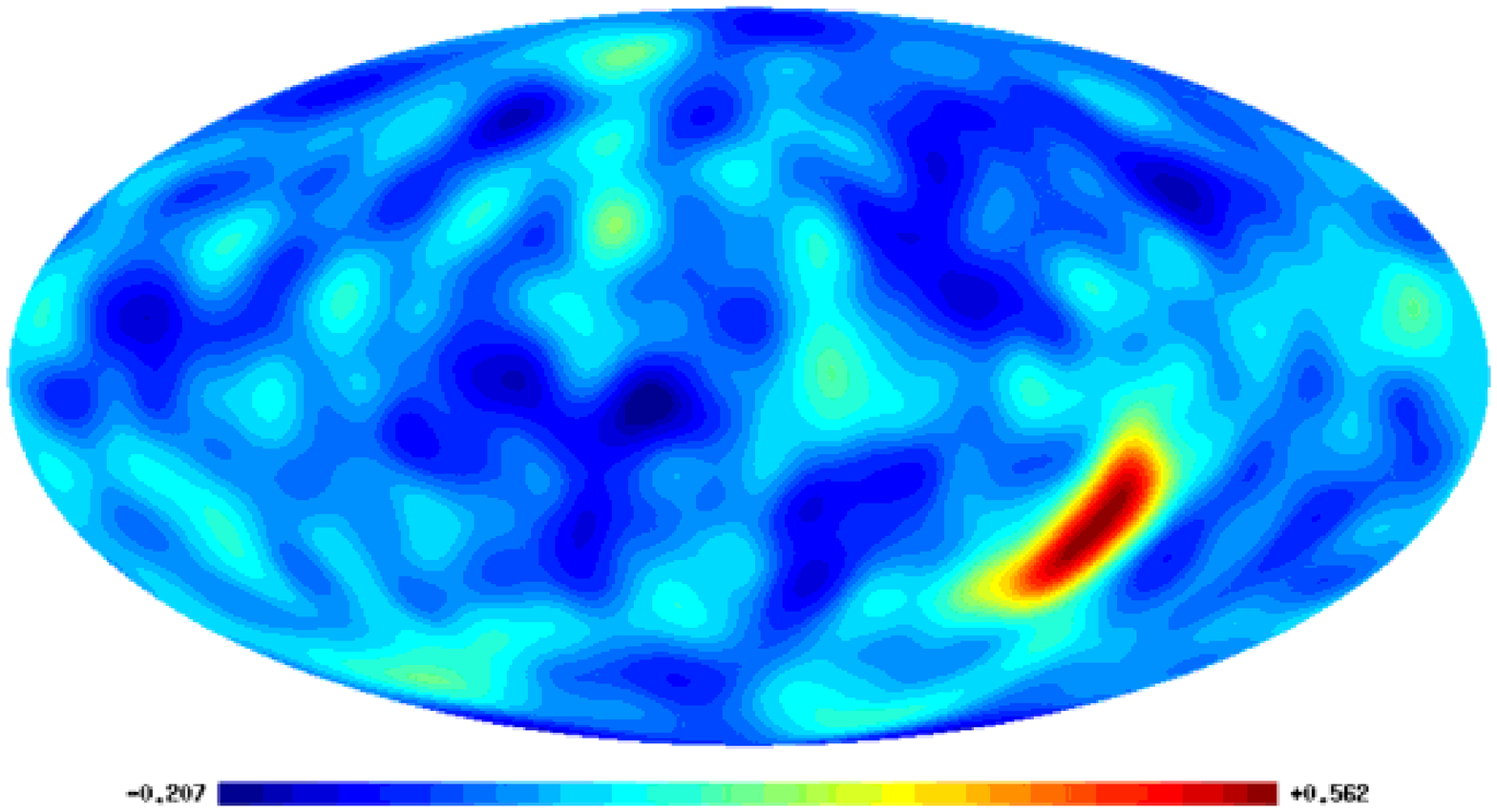}
\includegraphics[width=0.49\textwidth]{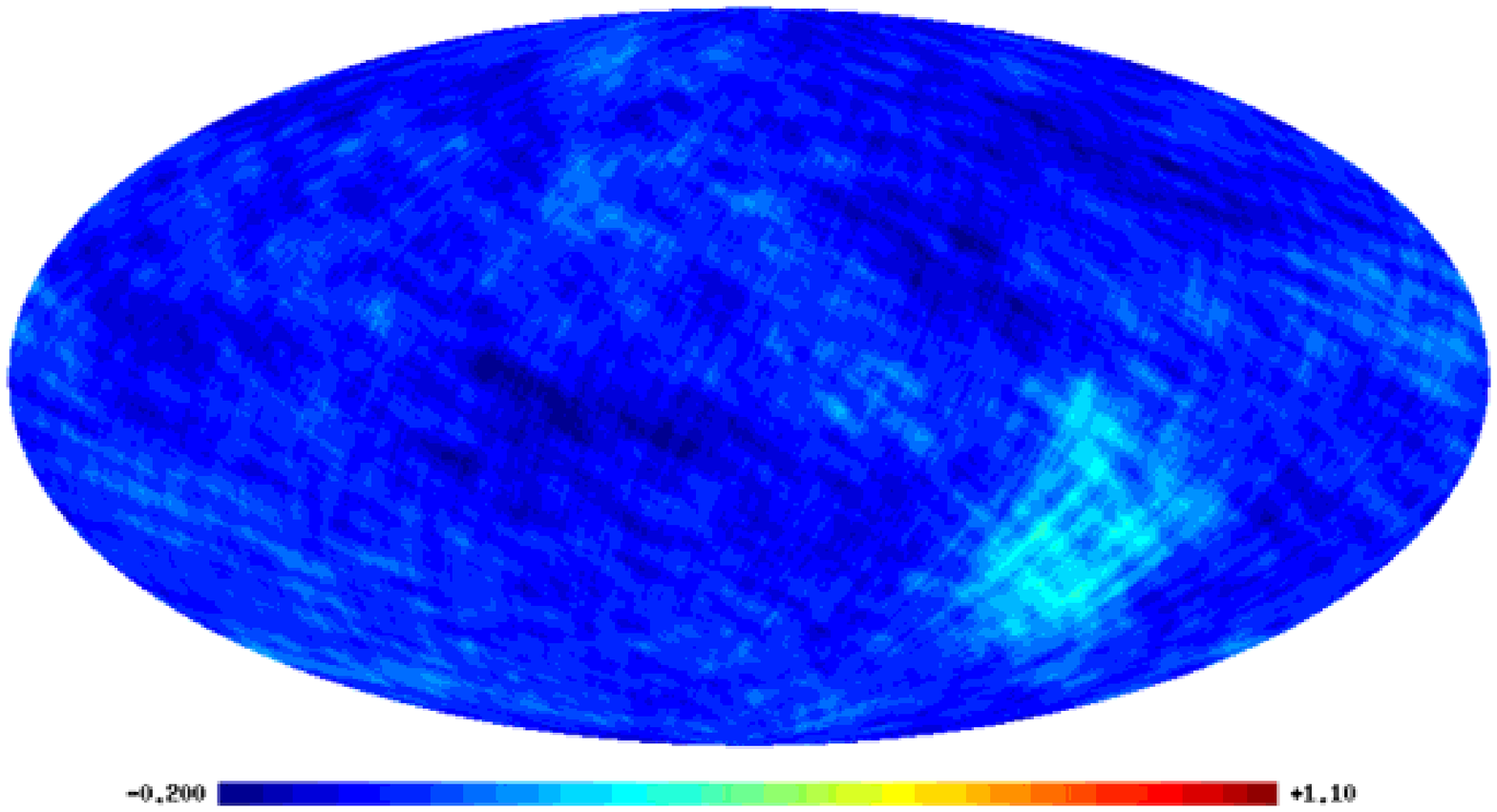} \hfill
\includegraphics[width=0.49\textwidth]{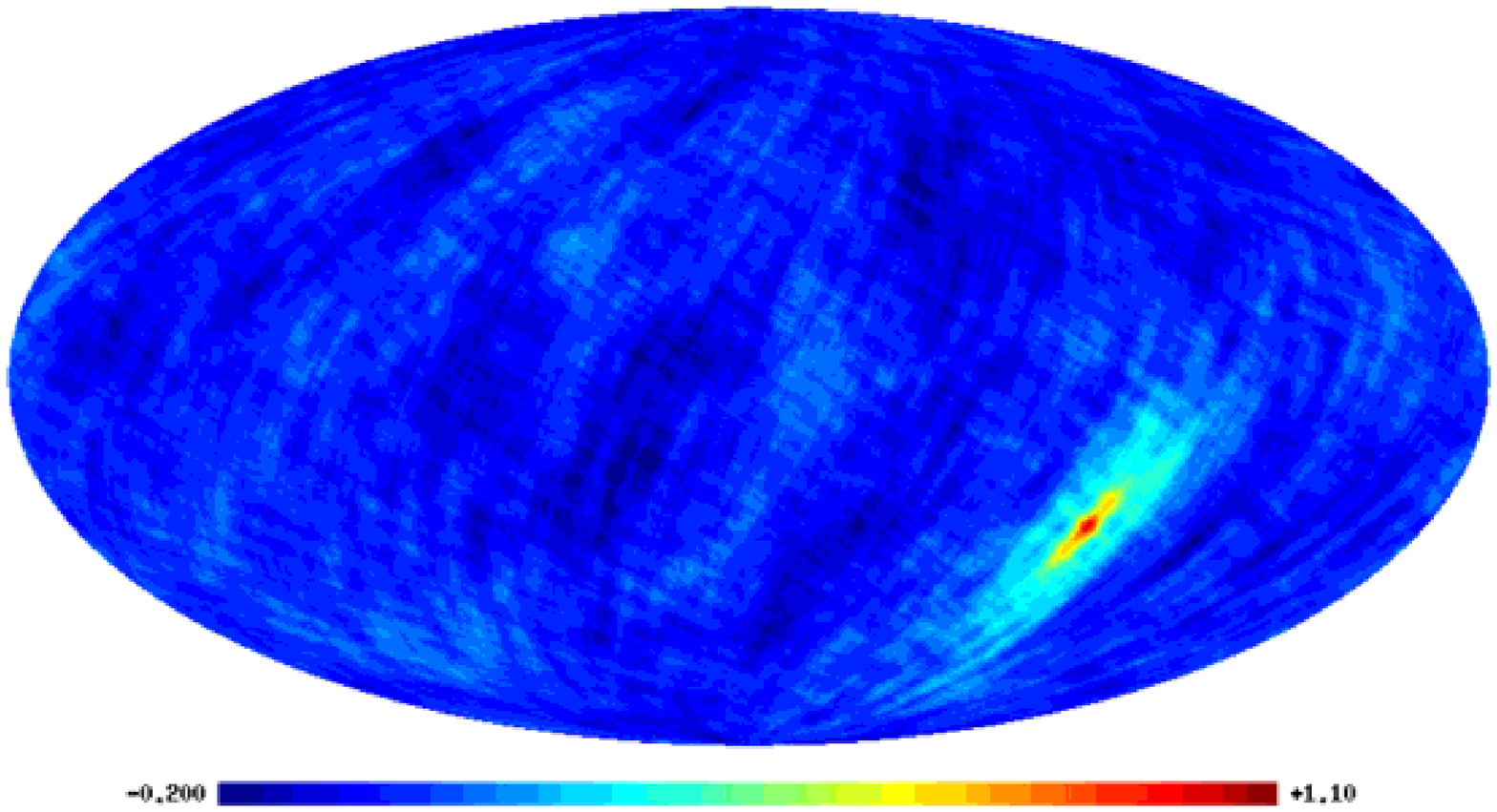}
\end{center}
\caption{
An example of an orientable matched filter.
(Top left) An orientable signal template, shown in Gnomonic projection.  The long dimension is $\sim 45^\circ$ and the short dimension is $\sim 13.5^\circ$.
(Top right) A signal, corresponding to the template, placed at $(\phi,\theta,\omega) = (270^\circ,120^\circ,210^\circ)$.
(Middle left) Signal plus white noise, note the change in color scale.
(Middle right) Signal plus noise, smoothed with a $13.5^\circ$ FWHM Gaussian.
(Bottom left)  Slice through the convolution, at an orientation $90^\circ$ from the true orientation.
(Bottom right) Slice through the convolution, at the true orientation.
} \label{fig:orientable_mf}
\end{figure*}

\begin{figure*}
\includegraphics[width=0.49\textwidth]{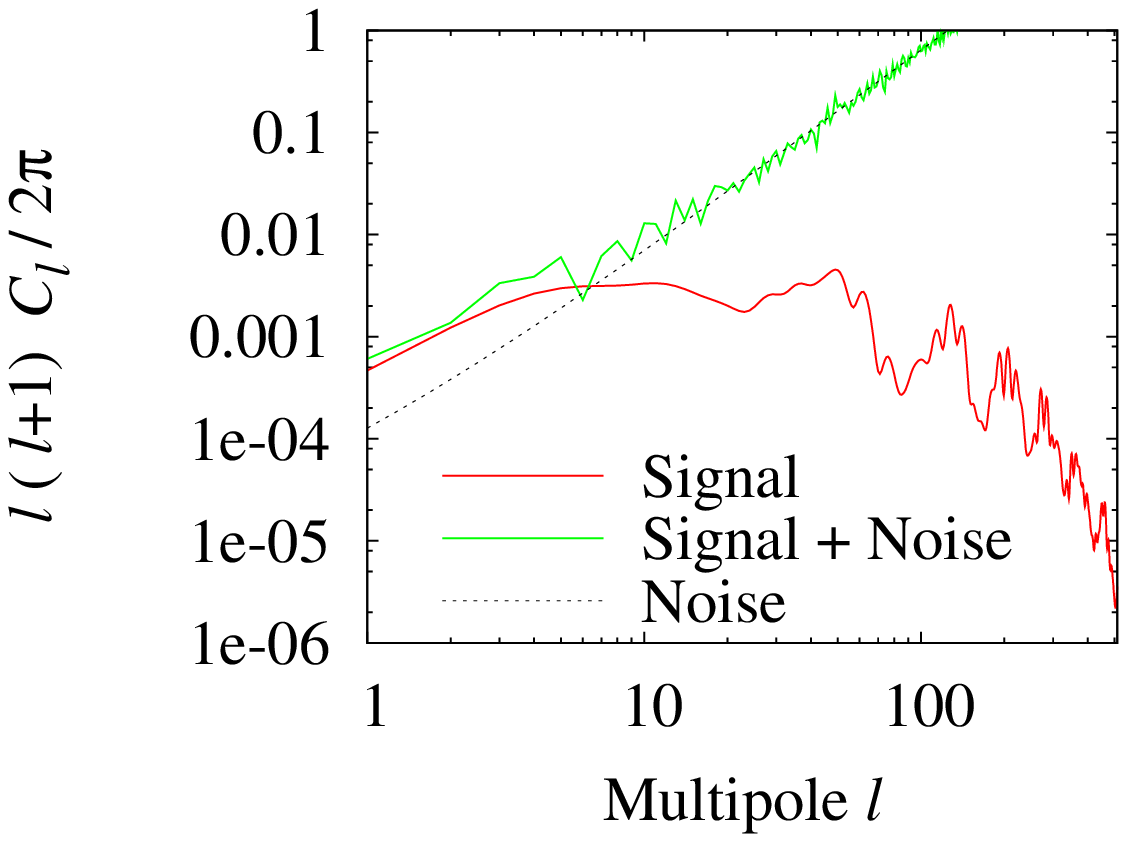} \hfill
\includegraphics[width=0.49\textwidth]{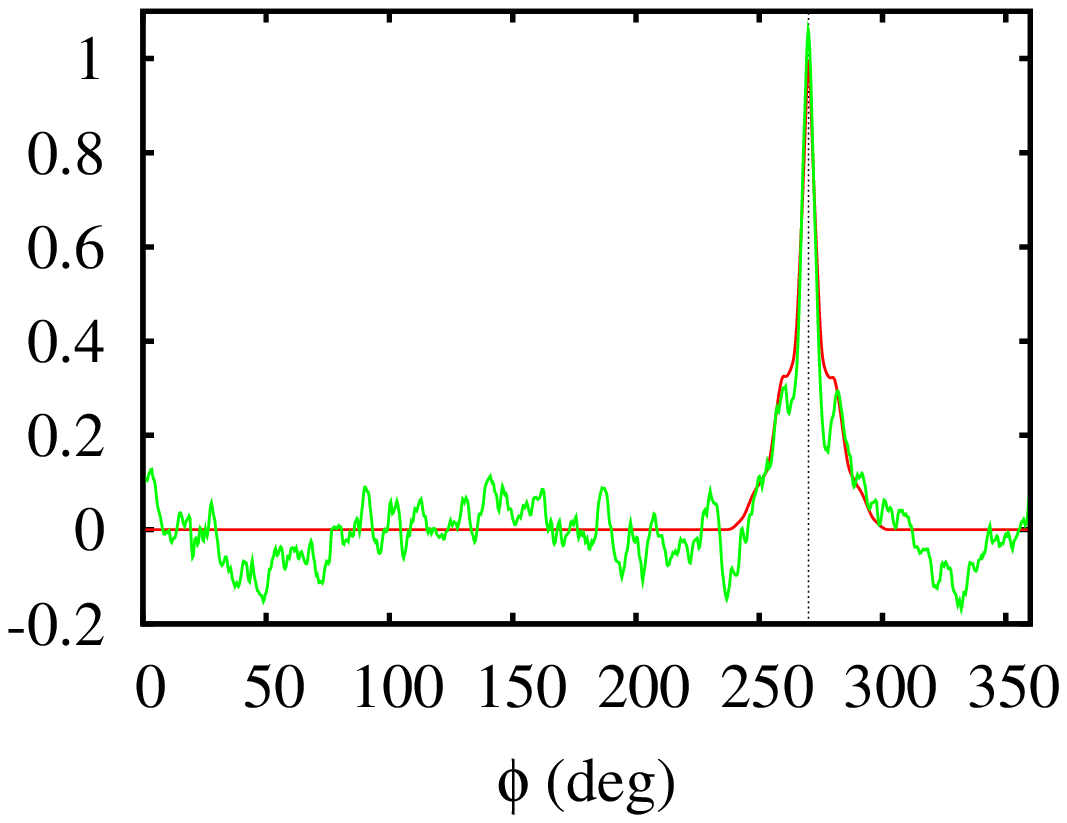}   \hfill
\includegraphics[width=0.49\textwidth]{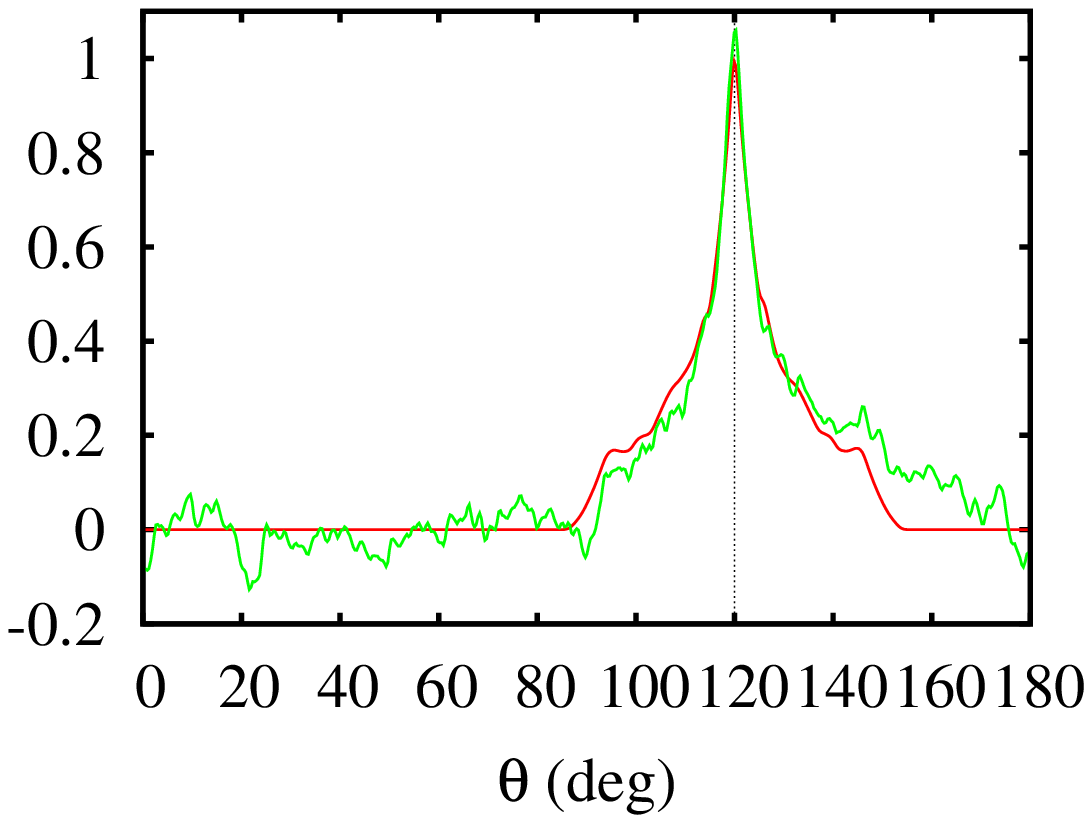} \hfill
\includegraphics[width=0.49\textwidth]{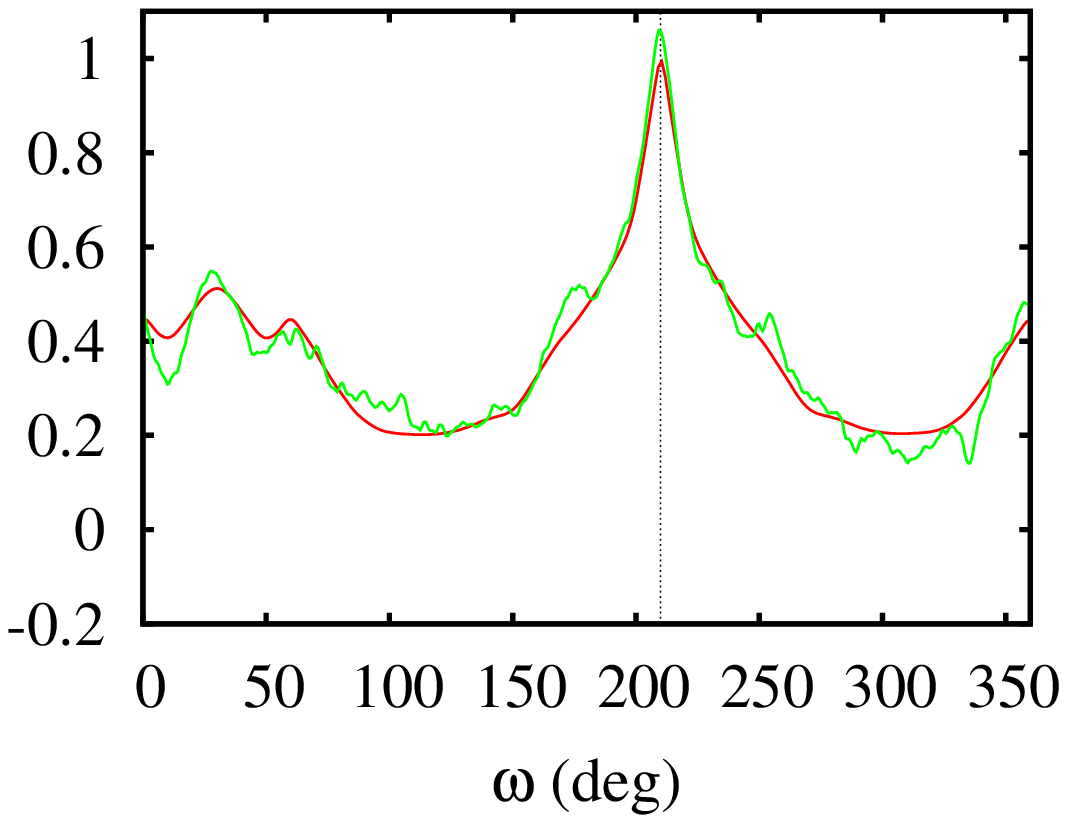} \hfill
\caption{Signal and noise power spectra for the case in figure \ref{fig:orientable_mf} (top-left), and slices through the convolution fixing all angles but one.  These slices pass through the maximum of the convolution, and show the recovery of the position and orientation of the signal at $(\phi,\theta,\omega) = (270^\circ,120^\circ,210^\circ)$.  Red is for signal alone; green includes noise.}
\label{fig:orientable_mf_graphs}
\end{figure*}

In figures \ref{fig:orientable_mf} and \ref{fig:orientable_mf_graphs}, we illustrate an example of the orientable matched filter.  We have hidden a beam with a nontrivial shape under white noise.  Only at the largest harmonic scales does the signal exceed the noise.  For this reason, a simple large Gaussian filter can discover the general location of the signal, but orientable matching filter finds it with much better precision, and can identify the orientation to high significance.

\begin{figure}
\end{figure}

\section{Discussion} \label{sec:discussion}

We have explored three applications of the spherical fast convolution formalism of \citet{WG2001}.  First, we examined the sidelobe pickup from \textit{CALISTO}, quantifying the level of contamination in three optical designs from galactic duct emission at 100 microns.  We found that the off-axis design with a cold stop permitted only a low level of contamination away from the galactic plane.   In future we wish to address the issue of sidelobe pickup from zodiacal light. We note that the sidelobe techniques described here are not at all specific to \textit{CALISTO}, and may be applied to any experiment.  In particular we are interested in further applications of this technique to the upcoming \textit{Planck} mission when detailed sidelobe beam maps become available.

Second, we examined the impact of asymmetric beams on scanning experiments, showing the difference in the smoothing of the CMB power spectrum as measured by \textit{Planck} from scanning with symmetric or asymmetric idealized beams.  Characterizing the beam asymmetries, and the impact on the maps, power spectrum, and associated errors, looks to be one of the most challenging data analysis problems for the \textit{Planck} mission.  The type of work displayed here is only a first step in approaching this problem.  Correcting this beam asymmetry effect remains an open research issue.

Third, we looked at orientable matched filters, showing that the use of orientation information can greatly aid the recovery of specific templates from noisy data.  We believe these techniques can find use in characterizing directional asymmetries and to look for systematics for all-sky surveys.

Finally, we believe that the fast convolution formalism is---and will continue to be---a useful tool for many data analysis applications in spherical geometry.  At this time, an optimized software implementation is being readied for public release.

\section*{Acknowledgments}
We thank Mark Dragovan and Paul Goldsmith for their suggestion to examine the \textit{CALISTO} beams, and for providing the beam models, useful discussion and feedback.  Charles Lawrence encouraged the \textit{CALISTO} work and provided useful comments on drafts of this paper.
We  acknowledge use of HEALPix software \citep{Gorski2005}, which helped produce many of the results in this work.  We thank Martin Reinecke for making his spherical harmonic transform library available.
IJO was supported by an appointment to the NASA Postdoctoral Program at 
the Jet Propulsion Laboratory, administered by Oak 
Ridge Associated Universities through a contract with NASA.
BDW acknowledges the Friedrich Wilhelm Bessel research award by the Alexander von Humboldt foundation and thanks the
Max Planck Institute for Astrophysics for hospitality. BDW is supported in part by NSF grant numbers AST 0507676 and 0708849, NASA/JPL subcontract no.~1236748. This research was supported in part by the National Science Foundation through TeraGrid resources provided by NCSA.
This work was partially performed at the Jet
Propulsion Laboratory, California Institute of Technology, under a
contract with NASA.

\bibliographystyle{apj}
\bibliography{apj-jour,tc_app}

\begin{thebibliography}{31}
\expandafter\ifx\csname natexlab\endcsname\relax\def\natexlab#1{#1}\fi

\bibitem[{{Ashdown {{\it et al}}.}(2007)}]{Trieste}
{Ashdown {{\it et al}}.}, M.~A.~J. 2007, in prep.

\bibitem[{{Barreiro} {et~al.}(2006){Barreiro}, {Mart{\'{\i}}nez-Gonz{\'a}lez},
  {Vielva}, \& {Hobson}}]{2006MNRAS.368..226B}
{Barreiro}, R.~B., {Mart{\'{\i}}nez-Gonz{\'a}lez}, E., {Vielva}, P., \&
  {Hobson}, M.~P. 2006, \mnras, 368, 226

\bibitem[{{Cay{\'o}n} {et~al.}(2006){Cay{\'o}n}, {Banday}, {Jaffe}, {Eriksen},
  {Hansen}, {Gorski}, \& {Jin}}]{2006MNRAS.369..598C}
{Cay{\'o}n}, L., {Banday}, A.~J., {Jaffe}, T., {Eriksen}, H.~K., {Hansen},
  F.~K., {Gorski}, K.~M., \& {Jin}, J. 2006, \mnras, 369, 598

\bibitem[{{Challinor} {et~al.}(2000){Challinor}, {Fosalba}, {Mortlock},
  {Ashdown}, {Wandelt}, \& {G{\'o}rski}}]{Challinor2000}
{Challinor}, A., {Fosalba}, P., {Mortlock}, D., {Ashdown}, M., {Wandelt}, B.,
  \& {G{\'o}rski}, K. 2000, \prd, 62, 123002

\bibitem[{{Cruz} {et~al.}(2006){Cruz}, {Tucci}, {Mart{\'{\i}}nez-Gonz{\'a}lez},
  \& {Vielva}}]{2006MNRAS.369...57C}
{Cruz}, M., {Tucci}, M., {Mart{\'{\i}}nez-Gonz{\'a}lez}, E., \& {Vielva}, P.
  2006, \mnras, 369, 57

\bibitem[{Edmonds(1960)}]{Edmonds}
Edmonds, A.~R. 1960, Angular momentum in quantum mechanics (Princeton, New
  Jersey: Princeton University Press)

\bibitem[{{Fosalba} {et~al.}(2002){Fosalba}, {Dor{\'e}}, \&
  {Bouchet}}]{Fosalba2002}
{Fosalba}, P., {Dor{\'e}}, O., \& {Bouchet}, F.~R. 2002, \prd, 65, 063003

\bibitem[{{Goldsmith} {et~al.}(2006){Goldsmith}, {Khayatian}, {Bradford},
  {Dragovan}, {Hoppe}, {Imbriale}, {Lee}, {Paine}, {Turner}, {Yorke}, \&
  {Zmuidzinas}}]{Goldsmith2006}
{Goldsmith}, P., {Khayatian}, B., {Bradford}, M., {Dragovan}, M., {Hoppe}, D.,
  {Imbriale}, W., {Lee}, R., {Paine}, C., {Turner}, P., {Yorke}, H., \&
  {Zmuidzinas}, J. 2006, in Proc. SPIE, Vol. 6265, Space Telescopes and
  Instrumentation I: Optical, Infrared, and Millimeter

\bibitem[{Goldsmith {et~al.}(2007)Goldsmith, Bradford, Dragovan, Khayatian,
  Huffenberger, O'Dwyer, Yorke, Zmuidzinas, Paine, Satter, \&
  Lee}]{Goldsmith2007}
Goldsmith, P.~F., Bradford, C.~M., Dragovan, M., Khayatian, B., Huffenberger,
  K., O'Dwyer, I., Yorke, H., Zmuidzinas, J., Paine, C., Satter, C., \& Lee, R.
  2007, in Proc. SPIE, in press

\bibitem[{{G{\'o}rski} {et~al.}(2005){G{\'o}rski}, {Hivon}, {Banday},
  {Wandelt}, {Hansen}, {Reinecke}, \& {Bartelmann}}]{Gorski2005}
{G{\'o}rski}, K.~M., {Hivon}, E., {Banday}, A.~J., {Wandelt}, B.~D., {Hansen},
  F.~K., {Reinecke}, M., \& {Bartelmann}, M. 2005, \apj, 622, 759

\bibitem[{{Hinshaw} {et~al.}(2007){Hinshaw}, {Nolta}, {Bennett}, {Bean},
  {Dor{\'e}}, {Greason}, {Halpern}, {Hill}, {Jarosik}, {Kogut}, {Komatsu},
  {Limon}, {Odegard}, {Meyer}, {Page}, {Peiris}, {Spergel}, {Tucker}, {Verde},
  {Weiland}, {Wollack}, \& {Wright}}]{Hinshaw2007}
{Hinshaw}, G., {Nolta}, M.~R., {Bennett}, C.~L., {Bean}, R., {Dor{\'e}}, O.,
  {Greason}, M.~R., {Halpern}, M., {Hill}, R.~S., {Jarosik}, N., {Kogut}, A.,
  {Komatsu}, E., {Limon}, M., {Odegard}, N., {Meyer}, S.~S., {Page}, L.,
  {Peiris}, H.~V., {Spergel}, D.~N., {Tucker}, G.~S., {Verde}, L., {Weiland},
  J.~L., {Wollack}, E., \& {Wright}, E.~L. 2007, \apjs, 170, 288

\bibitem[{{Jaffe} {et~al.}(2005){Jaffe}, {Banday}, {Eriksen}, {G{\'o}rski}, \&
  {Hansen}}]{2005ApJ...629L...1J}
{Jaffe}, T.~R., {Banday}, A.~J., {Eriksen}, H.~K., {G{\'o}rski}, K.~M., \&
  {Hansen}, F.~K. 2005, \apjl, 629, L1

\bibitem[{{Jaffe} {et~al.}(2006{\natexlab{a}}){Jaffe}, {Banday}, {Eriksen},
  {G{\'o}rski}, \& {Hansen}}]{2006A&A...460..393J}
---. 2006{\natexlab{a}}, \aap, 460, 393

\bibitem[{{Jaffe} {et~al.}(2006{\natexlab{b}}){Jaffe}, {Banday}, {Eriksen},
  {G{\'o}rski}, \& {Hansen}}]{2006ApJ...643..616J}
---. 2006{\natexlab{b}}, \apj, 643, 616

\bibitem[{{Liu} \& {Zhang}(2006)}]{2006ApJ...636L...1L}
{Liu}, X., \& {Zhang}, S.-N. 2006, \apjl, 636, L1

\bibitem[{{McEwen} {et~al.}(2006){McEwen}, {Hobson}, {Lasenby}, \&
  {Mortlock}}]{2006MNRAS.369.1858M}
{McEwen}, J.~D., {Hobson}, M.~P., {Lasenby}, A.~N., \& {Mortlock}, D.~J. 2006,
  \mnras, 369, 1858

\bibitem[{{McEwen} {et~al.}(2007{\natexlab{a}}){McEwen}, {Vielva}, {Wiaux},
  {Barreiro}, {Cayon}, {Hobson}, {Lasenby}, {Mart{\'{\i}}nez-Gonz{\'a}lez}, \&
  {Sanz}}]{2007arXiv0704.3158M}
{McEwen}, J.~D., {Vielva}, P., {Wiaux}, Y., {Barreiro}, R.~B., {Cayon}, L.,
  {Hobson}, M.~P., {Lasenby}, A.~N., {Mart{\'{\i}}nez-Gonz{\'a}lez}, E., \&
  {Sanz}, J.~L. 2007{\natexlab{a}}, ArXiv e-prints, 704

\bibitem[{{McEwen} {et~al.}(2007{\natexlab{b}}){McEwen}, {Wiaux}, {Hobson},
  {Vandergheynst}, \& {Lasenby}}]{2007arXiv0704.0626M}
{McEwen}, J.~D., {Wiaux}, Y., {Hobson}, M.~P., {Vandergheynst}, P., \&
  {Lasenby}, A.~N. 2007{\natexlab{b}}, ArXiv e-prints, 704

\bibitem[{{Page} {et~al.}(2003){Page}, {Barnes}, {Hinshaw}, {Spergel},
  {Weiland}, {Wollack}, {Bennett}, {Halpern}, {Jarosik}, {Kogut}, {Limon},
  {Meyer}, {Tucker}, \& {Wright}}]{2003ApJS..148...39P}
{Page}, L., {Barnes}, C., {Hinshaw}, G., {Spergel}, D.~N., {Weiland}, J.~L.,
  {Wollack}, E., {Bennett}, C.~L., {Halpern}, M., {Jarosik}, N., {Kogut}, A.,
  {Limon}, M., {Meyer}, S.~S., {Tucker}, G.~S., \& {Wright}, E.~L. 2003, \apjs,
  148, 39

\bibitem[{Reinecke(2006)}]{Reinecke2006personal}
Reinecke, M. 2006, personal communication

\bibitem[{{Reinecke} {et~al.}(2006){Reinecke}, {Dolag}, {Hell}, {Bartelmann},
  \& {En{\ss}lin}}]{LevelS}
{Reinecke}, M., {Dolag}, K., {Hell}, R., {Bartelmann}, M., \& {En{\ss}lin},
  T.~A. 2006, \aap, 445, 373

\bibitem[{Risbo(1996)}]{Risbo1996}
Risbo, T. 1996, Journal of Geodesy, 70, 383

\bibitem[{{Schlegel} {et~al.}(1998){Schlegel}, {Finkbeiner}, \&
  {Davis}}]{SFD1998}
{Schlegel}, D.~J., {Finkbeiner}, D.~P., \& {Davis}, M. 1998, \apj, 500, 525

\bibitem[{{The Planck Collaboration}(2006)}]{PlanckBluebook}
{The Planck Collaboration}. 2006, ArXiv Astrophysics e-prints

\bibitem[{{Vielva} {et~al.}(2006{\natexlab{a}}){Vielva},
  {Mart{\'{\i}}nez-Gonz{\'a}lez}, \& {Tucci}}]{2006MNRAS.365..891V}
{Vielva}, P., {Mart{\'{\i}}nez-Gonz{\'a}lez}, E., \& {Tucci}, M.
  2006{\natexlab{a}}, \mnras, 365, 891

\bibitem[{{Vielva} {et~al.}(2006{\natexlab{b}}){Vielva}, {Wiaux},
  {Mart{\'{\i}}nez-Gonz{\'a}lez}, \& {Vandergheynst}}]{2006NewAR..50..880V}
{Vielva}, P., {Wiaux}, Y., {Mart{\'{\i}}nez-Gonz{\'a}lez}, E., \&
  {Vandergheynst}, P. 2006{\natexlab{b}}, New Astronomy Review, 50, 880

\bibitem[{{Vielva} {et~al.}(2007){Vielva}, {Wiaux},
  {Mart{\'{\i}}nez-Gonz{\'a}lez}, \& {Vandergheynst}}]{2007arXiv0704.3736V}
---. 2007, ArXiv e-prints, 704

\bibitem[{{Wandelt} \& {G{\'o}rski}(2001)}]{WG2001}
{Wandelt}, B.~D., \& {G{\'o}rski}, K.~M. 2001, \prd, 63, 123002

\bibitem[{{Wiaux} {et~al.}(2007{\natexlab{a}}){Wiaux}, {McEwen}, \&
  {Vielva}}]{2007arXiv0704.3144W}
{Wiaux}, Y., {McEwen}, J.~D., \& {Vielva}, P. 2007{\natexlab{a}}, ArXiv
  e-prints, 704

\bibitem[{{Wiaux} {et~al.}(2007{\natexlab{b}}){Wiaux}, {Vielva}, {Barreiro},
  {Mart{\'{\i}}nez-Gonz{\'a}lez}, \& {Vandergheynst}}]{2007arXiv0706.2346W}
{Wiaux}, Y., {Vielva}, P., {Barreiro}, R.~B., {Mart{\'{\i}}nez-Gonz{\'a}lez},
  E., \& {Vandergheynst}, P. 2007{\natexlab{b}}, ArXiv e-prints, 706

\bibitem[{{Wiaux} {et~al.}(2006){Wiaux}, {Vielva},
  {Mart{\'{\i}}nez-Gonz{\'a}lez}, \& {Vandergheynst}}]{2006PhRvL..96o1303W}
{Wiaux}, Y., {Vielva}, P., {Mart{\'{\i}}nez-Gonz{\'a}lez}, E., \&
  {Vandergheynst}, P. 2006, Physical Review Letters, 96, 151303

\end{thebibliography}

\end{document}